\documentclass[twocolumn]{aastex631} 

\begin{document}

\title{Design and Implementation of a Scalable Correlator Based on ROACH2+GPU Cluster for Tianlai 96-Dual-Polarization Antenna Array}

\author[0009-0007-8230-9798]{Zhao Wang}
\affiliation{Center for Astronomy and Space Sciences, China Three Gorges University, Yichang 443002, China}

\author[0000-0001-9652-1377]{Ji-Xia Li}
\affiliation{National Astronomical Observatories, Chinese Academy of Sciences, Beijing 100101, China}

\author{Ke Zhang}
\affiliation{Center for Astronomy and Space Sciences, China Three Gorges University, Yichang 443002, China}

 \author[0000-0002-6174-8640]{Feng-Quan Wu}
\affiliation{National Astronomical Observatories, Chinese Academy of Sciences, Beijing 100101, China}

  \author[0000-0001-9289-0589]{Hai-Jun Tian}\thanks{Corresponding Emails: jxli@bao.ac.cn, wufq@bao.ac.cn, \\hjtian@hdu.edu.cn, zhangjy@hdu.edu.cn, chen\_zp@hdu.edu.cn.}
\affiliation{School of Science, Hangzhou Dianzi University, Hangzhou, 310018, China}
\affiliation{Big Data Institute, Hangzhou Dianzi University, Hangzhou, 310018, China}

   \author[0000-0001-6651-7799]{Chen-Hui Niu}
\affiliation{Central Normal University, Wuhan, 100101, China}

   \author{Ju-Yong Zhang}
\affiliation{School of Science, Hangzhou Dianzi University, Hangzhou, 310018, China}
\affiliation{Big Data Institute, Hangzhou Dianzi University, Hangzhou, 310018, China}

  \author{Zhi-Ping Chen}
\affiliation{School of Science, Hangzhou Dianzi University, Hangzhou, 310018, China}
\affiliation{Big Data Institute, Hangzhou Dianzi University, Hangzhou, 310018, China}

  \author{Dong-Jin Yu}
\affiliation{Big Data Institute, Hangzhou Dianzi University, Hangzhou, 310018, China}

 \author[0000-0001-6475-8863]{Xue-Lei Chen}
\affiliation{National Astronomical Observatories, Chinese Academy of Sciences, Beijing 100101, China}







\begin{abstract}
The digital correlator is one of the most crucial data processing components of a radio telescope array. With the scale of radio interferometeric array growing, many efforts have been devoted to developing a cost-effective and scalable correlator in the field of radio astronomy. In this paper,
a 192-input digital correlator with six CASPER ROACH2 boards and seven GPU servers has been deployed as the digital signal processing system for Tianlai cylinder pathfinder located in Hongliuxia observatory.
The correlator consists of 192 input signals (96 dual-polarization), 125-MHz bandwidth, and full-Stokes output. The correlator inherits the advantages of the CASPER system, for example, low cost, high performance, modular scalability, and a heterogeneous computing architecture. With a rapidly deployable ROACH2 digital sampling system, a commercially  expandable 10 Gigabit switching network system, and a flexible upgradable GPU computing system, the correlator forms a low-cost and easily-upgradable system, poised to support scalable large-scale interferometeric array in the future.
\end{abstract}

\keywords{Techniques: interferometric; instrumentation: interferometers}


\section{Introduction} \label{sec:intro}
The digital correlator plays a crucial role in radio astronomy by combining individual antennas to form a large-aperture antenna, keeping large field of view, and providing high-resolution images. 
At present, many radio interferometric
arrays in the world use CASPER (Collaboration for Astronomy Signal Processing and Electronics Research) hardware platform ROACH2 (Reconfigurable Open Architecture Computing Hardware-2) to develop correlators. For example, PAPER (Precision Array for Probing the Epoch of Reionization)  in South Africa's Karoo Desert \citep{PAPER,PAPER64}. The 100 MHz FX correlator was originally based on iBOBs (Interconnect Break-out Boards) and later upgraded to ROACH, and then ROACH2 boards \citep{CASPER-Decade-Develop}. Currently, PAPER uses 8 ROACH2 boards for channelization, followed by a GPU (Graphics Processing Unit)-based `X' stage. Additionally, the `large-N' correlator located in the Owens Valley Radio Observatory (LWA-OV) is designed to enable the Large Aperture Experiment to Detect the Dark Ages (LEDA) \citep{LEDA}. It features a 58 MHz, 512-input digitization, channelization, and packetization system using a GPU correlator backend. 

\begin{figure*}[ht]
\centering
\includegraphics[width=0.7\textwidth]{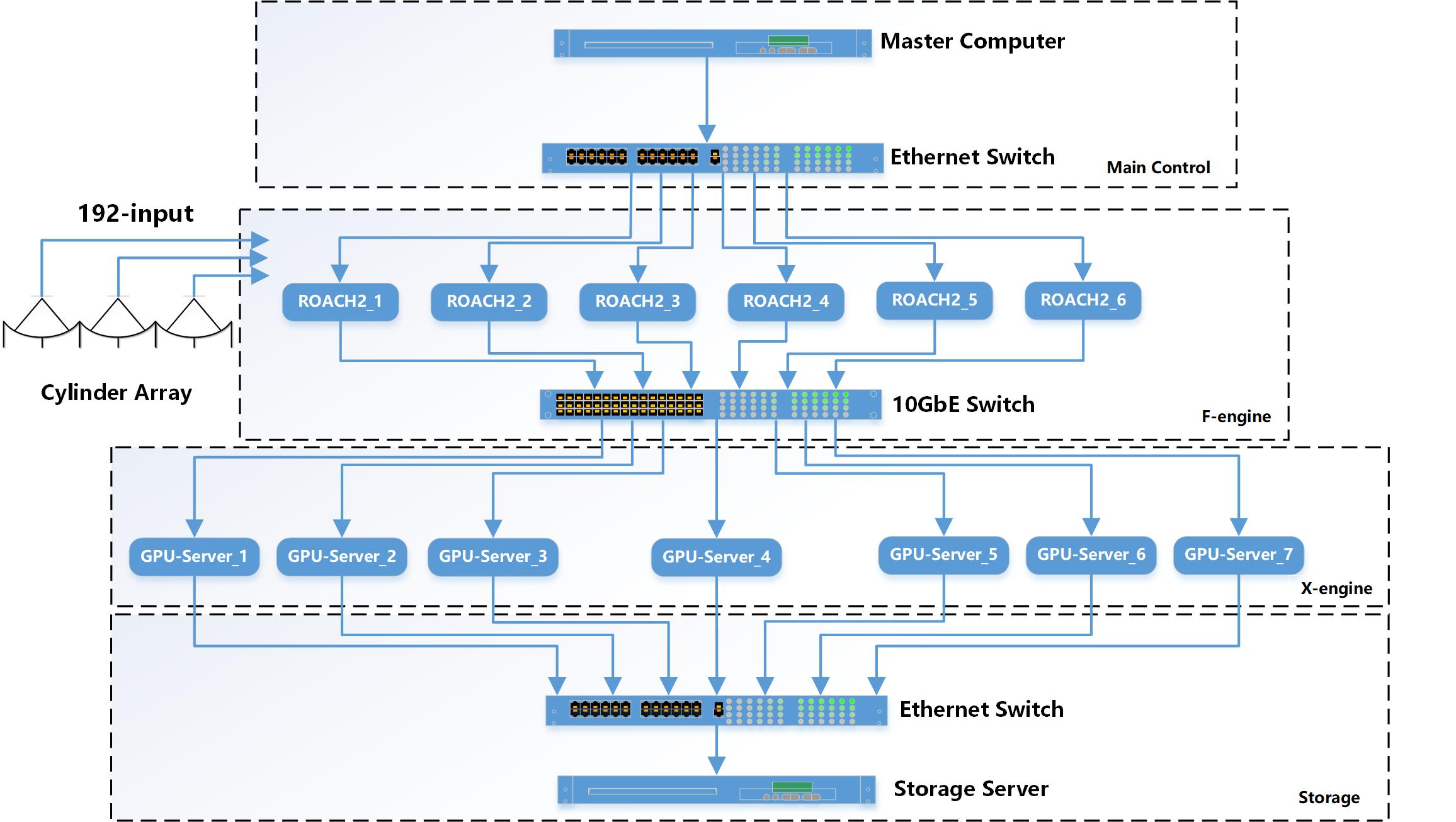}
\caption{This block diagram illustrates the Tianlai cylinder array correlator. The master computer communicates with ROACH2 boards, a 10\,GbE switch, and GPU servers through an Ethernet switch. Six ROACH2 boards receive 192 input signals from the antenna. After the signal processing is completed, the UDP(User Datagram Protocol) data is sent to the 10\,GbE switch. Seven GPU servers receive the UDP data to calculate the cross-correlation, sending results back to the Ethernet switch after the computations are finished. Finally, the data is transmitted to the master computer for storage via the Ethernet switch.
\label{fig:block_diagram}}
\end{figure*}

The Tianlai project \footnote{\url{https://tianlai.bao.ac.cn}} is an experiment aimed at detecting dark energy by measuring baryon acoustic oscillation (BAO) features in the large-scale structure power spectrum, in which BAO can be used as a standard ruler\citep{bao_lss_reza,intensity_mapping,21cm_AdrianLiu}. The basic plan is to build a radio telescope array and use it to make 21cm intensity mapping observations of neutral hydrogen, which trace the large-scale structure of the matter distribution \citep{xuelei2011-tianlai,tianlai_forecast_xuyd,zhangjiao_sky,tianlai_progress,yukf_apply,map_err_yukf}. Currently, two different types of pathfinder array have been built in a quiet radio site in Hongliuxia, Balikun county, Xinjiang, China \citep{site-selection}. The cylinder array consists of three adjacent parabolic cylinder reflectors, each 40m $\times$ 15m, with their long axes oriented in the N-S direction. It has a total of 96 dual-polarization feeds, resulting in 192 signal channels \citep{feed_test,lijixia2020-cyl,ssj_emsim}. 
The dish array includes 16 dishes with 6-meter aperture. Each dish has a dual polarization feed, generating 32 signal channels in total \citep{wufq2021-dish,dish_forecast,dish_couping}. 
In addition to the ability to survey the 21cm Hydrogen sky, both antenna arrays are also capable of detecting fast radio bursts \citep{dish_frb,dish_frb_telegram,cylinder_frb_telegram}. 
This paper is about the development of correlator for Tianlai cylinder pathfinder array.



The design of the Tianlai cylinder correlator is based on the prototype correlator of \citet{niuchenhui2019-roach}, which has 32 inputs and was used for the Tianlai Dish pathfinder array. This 32-input prototype correlator is built upon the model of PAPER correlator, which creates a flexible and scalable hybrid correlator system. We expanded the prototype correlator from 32 to 192 channels, reprogrammed the network transport model, increased it from a single GPU server to seven GPU servers, solved the synchronization problem of multiple devices. It was eventually deployed in the machine room on the Hongliuxia site. The primary motivation behind the design of the PAPER correlator architecture is the scalability for large-scale antenna arrays, and it has been executed exceptionally well. Therefore, we have chosen to borrow ideas from the PAPER correlator. The Tianlai project is expanding and the number of single inputs will soon increase to more than 500.

The Tianlai cylinder correlator is a flexible, scalable, and efficient system, which has a hybrid structure of ROACH2+GPU+10\,GbE network. A ROACH2 is an independent board, unlike a PCIe-sampling board which needs to be plugged into a computer server and often leads to some incompatible issues. A GPU card is dramatically upgrading, and it is almost the best choice among the current available hardwares, such as CPU/GPU/DSP(Digital Signal Processing), by comprehensively considering the flexibility, the efficiency and the cost. The module of the data switch network is easy to be upgraded, since the Ethernet switch has a variety of commercial applications. We have uploaded all the project files to Github. \footnote{https://github.com/TianlaiProject} 

This paper gives a detailed introduction to the function and performance of the Tianlai 192-input cylinder correlator system. In Section \ref{sec:sys_design}, we introduce the general framework of the correlator system and show the deployment of the correlator. Then, in a sub-section, we provide a detailed introduction to the design and functions of each module. In Section  \ref{sec:test}, we evaluate the performance of the correlator. Section \ref{sec:summary} summarizes the correlator system and presents the design scheme for correlator expansion in the future as part of the Tianlai project. 

\section {System design} \label{sec:sys_design}

The digital correlators can be classified into two types: XF and FX. XF correlators combine signals from multiple antennas and performs cross-correlation followed by Fourier transformation.  XF correlators can handle a large number of frequency channels and have a relatively simple hardware design \citep{thompson-book}. FX correlators combine signals from multiple antennas and perform the Fourier transformation followed by cross-correlation. FX correlators can handle a large number of antenna pairs and also have a relatively simple hardware design. The Tianlai cylinder correlator is an FX correlator. 

The Tianlai cylinder correlator system can be divided into four parts, as shown in Figure \ref{fig:block_diagram}. The first part is the control part which consists of a master computer and an Ethernet switch. The Ethernet switch is used for net-booting of the ROACH2, monitoring the status of F-engine and hashpipe \footnote{https://github.com/david-macmahon/hashpipe}, and synchronizing the running status of F-engine and X-engine.



The second part is the F-engine, which consists of six ROACH2 boards and one 10\,GbE switch. The 192 input signals from the Tianlai cylinder array are connected to the ADC connectors on the ROACH2 boards. The main functions of the F-engine are to Fourier transform the data from the time domain into the frequency domain, and transmit the data to the GPU server through a 10\,GbE switch.

The third part is the X-engine, which performs cross-correlation on the received Fourier data. Each GPU server receives packets from all ROACH2 boards. The details of network transmission will be explained later. The X-engine utilizes a software called hashpipe \citep{breakthrough_hashpipe} to store, deliver and compute the cross-correlations.

The fourth part is the data storage part, which consists of seven GPU servers, an Ethernet switch, and a storage server (shared with the master computer). The GPU servers transmit data to the storage server via an Ethernet switch. We have developed a multi-threading program to collect and organize data packets from different GPU servers, and finally save them  onto hard drives in HDF5 format.

\begin{figure}[ht]
\graphicspath{{./fig/}}
\centering
\rotatebox{0}{\includegraphics[width=0.33\textwidth]{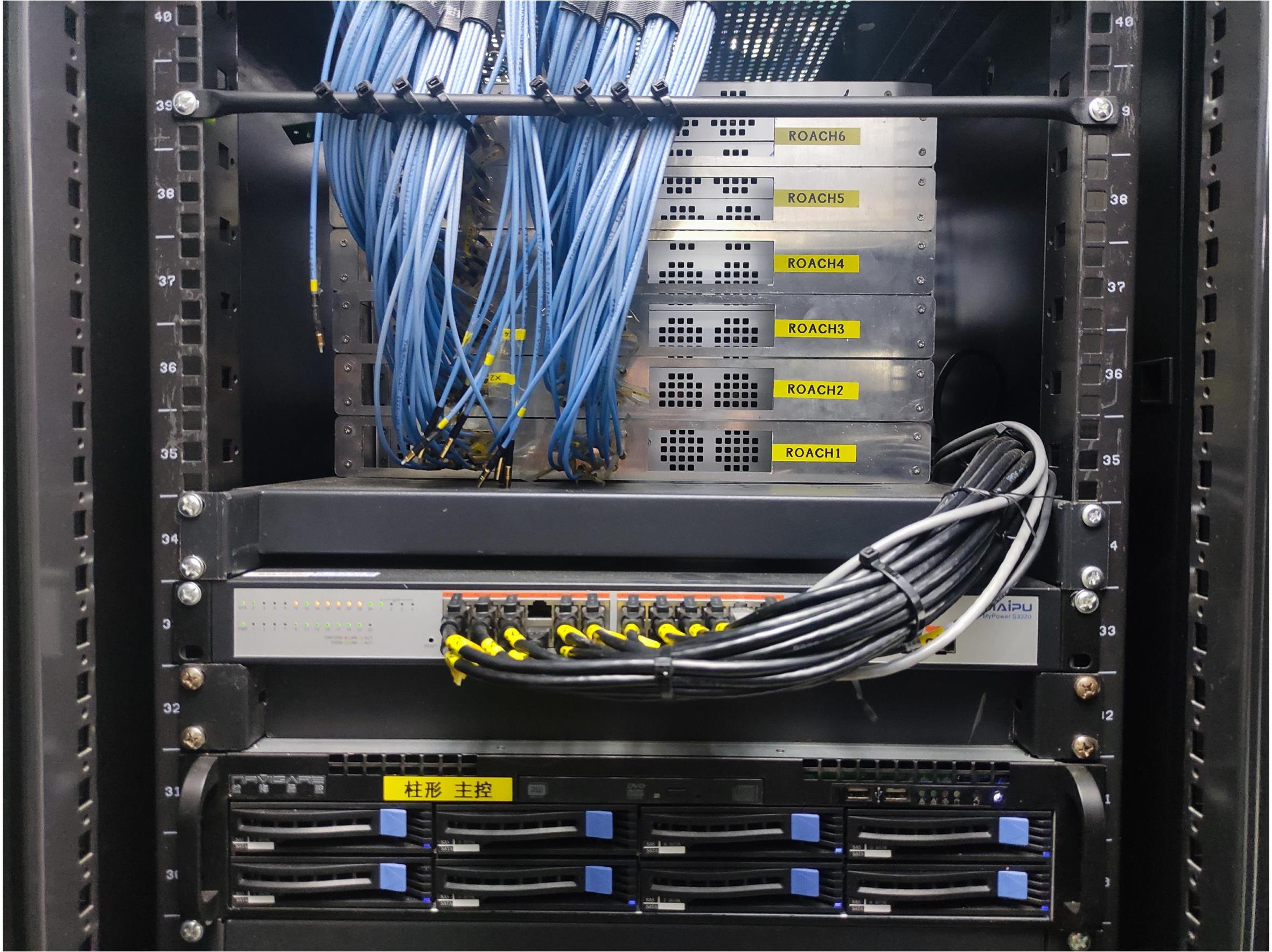}}

(a)

\rotatebox{0}{\includegraphics[width=0.33\textwidth]{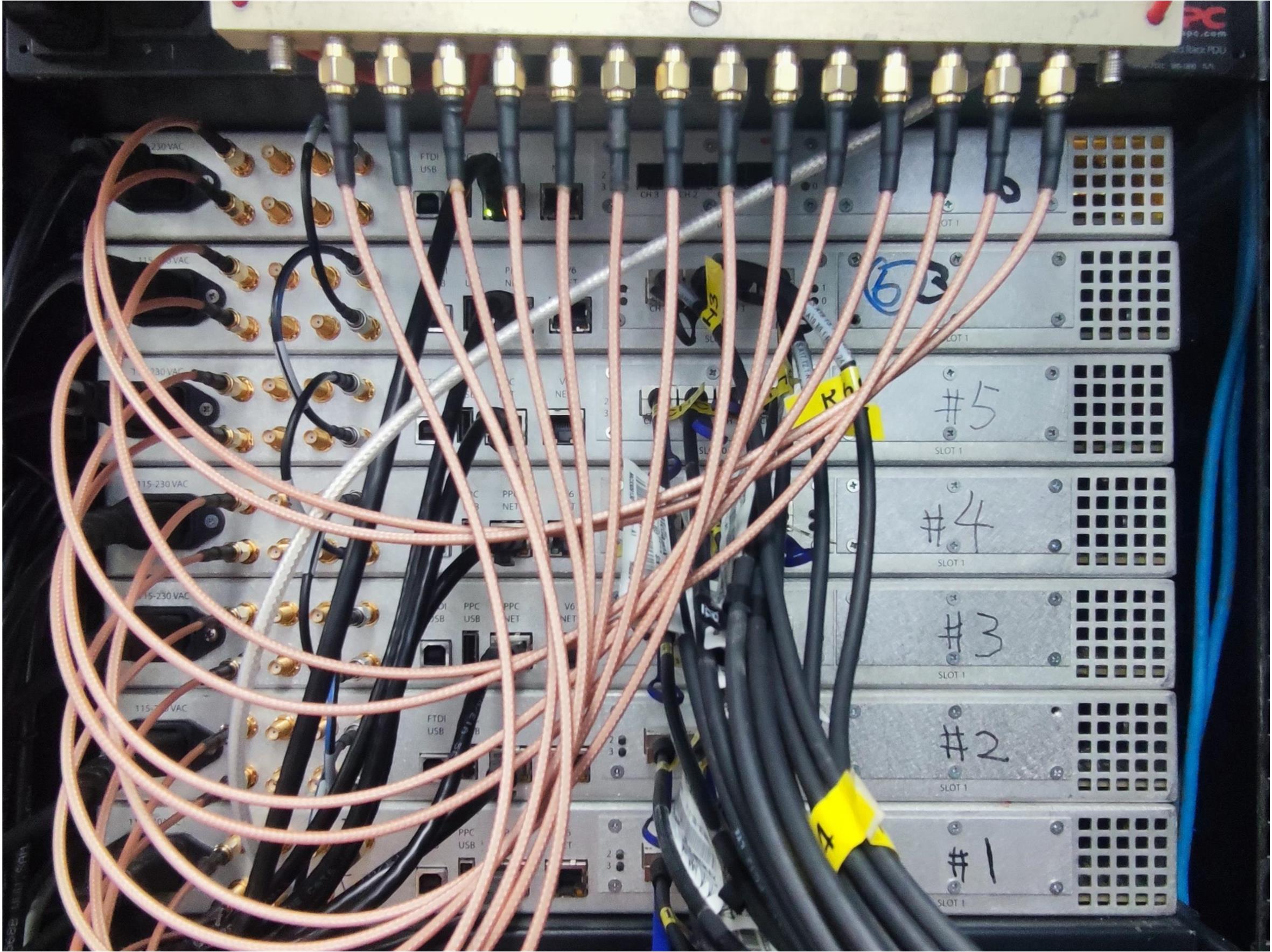}}

(b)

\rotatebox{0}{\includegraphics[width=0.33\textwidth]{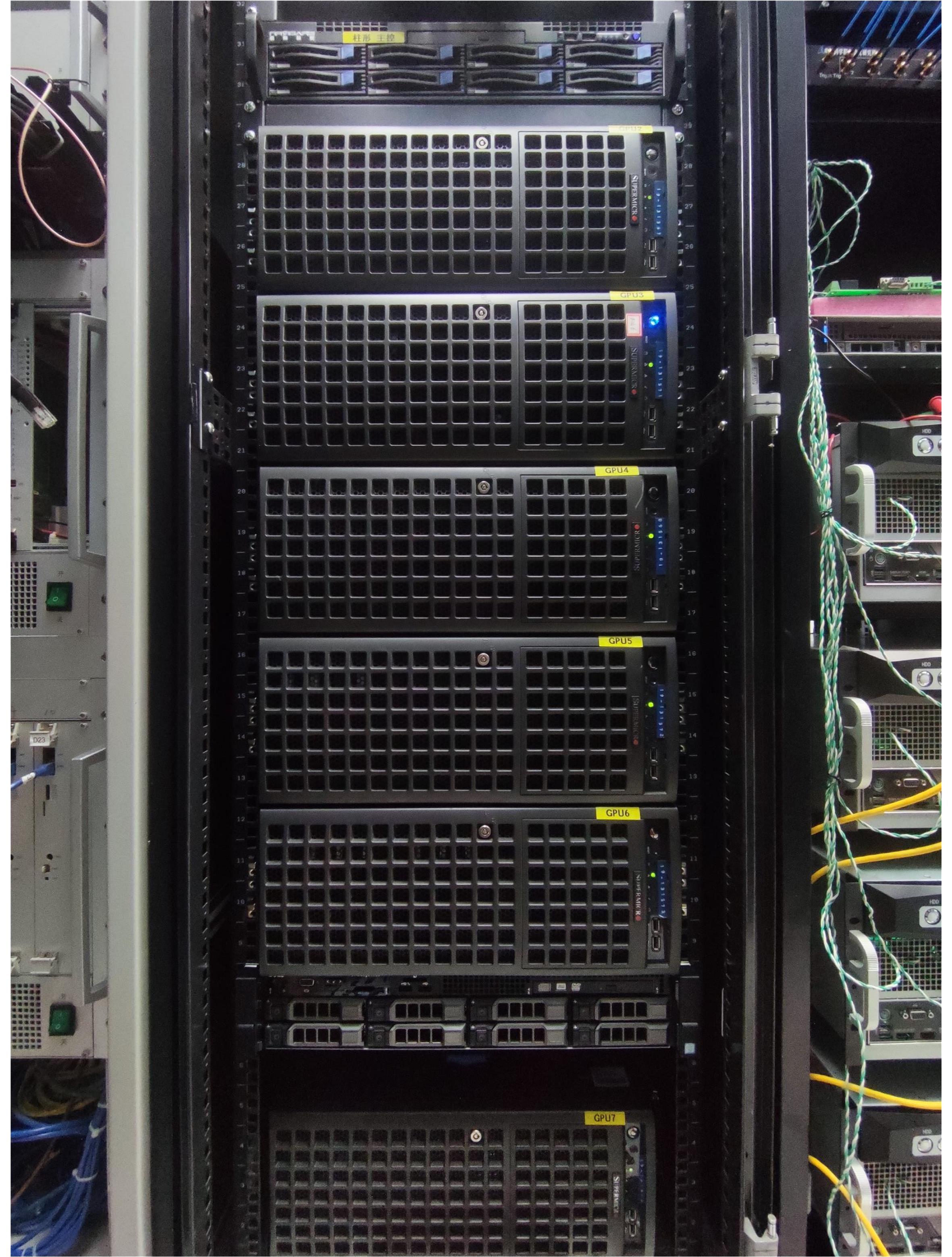}}

(c)
\caption{(a) Six ROACH2 boards are connected to 192 input signals. The two switches are positioned on the same layer, with the Ethernet switch in the front and the 10\,GbE switch at the rear. The master computer is located below.
(b) Clock and synchronization cable connections on the rear panel of ROACH2 boards. (c) Seven GPU servers are vertically stacked, consisting of six Supermicro servers with a size of 4U and a Dell server with a size of 2U.}
\label{fig:system_photo}
\end{figure}

The deployment of the correlator system is shown in Figure \ref{fig:system_photo}. It consists of six ROACH2 boards, an Ethernet switch, a 10\,GbE switch, a master computer, and seven GPU servers, arranged from top to bottom. The yellow ``ROACH2'' label in Figure \ref{fig:system_photo}(a) represents the front panel of the ROACH2 board. (In our case, a connector transformer panel has been specifically designed to conveniently connect to the radio cables.)
The ADC connector of the ROACH2 is connected to a blue RF (Radio Frequency) cable that transmits the analog signal. Figure \ref{fig:system_photo}(b) shows the back side of the ROACH2 board. On the far left is the power line. The light orange RF cable is the clock cable. The 250~MHz clock of the ROACH2 board is output by a VALON 5008 dual-frequency synthesizer module, and it is split by a 12-way power splitter. The short blue-black cable connects to the synchronization port between the ROACH2 boards. We use synchronization ports in F-engine functional block design to ensure that the six ROACH2 boards work at the same clock. The signal of the synchronization port is provided by a time server. The time server sends out a 1-PPS (Pulse Per Second) signal, which is used to initialize the synchronization module of the F-engine system. After running the F-engine control script, the 1 PPS signal  drives the F-engine and synchronizes the operational state of the six ROACH2 boards.
The bandwidth of the antenna signal input to the ROACH2 board is 125 MHz. According to the Nyquist sampling law, the input signal can be completely recovered by a 250 Msps sampling rate. In Figure \ref{fig:system_photo}(c), seven GPU servers are vertically stacked, consisting of six Supermicro servers with a size of 4U (Unit) and one Dell server with a size of 2U. These devices are used to implement X-engine functionality. The number of servers is determined by the total frequency channel count and the frequency channel processing capacity of each server. In terms of computational performance, each GPU server runs 4 hashpipe threads, processing a total of 128 frequency channels. At this configuration, the computational performance accounts for approximately 46\% of the theoretical peak performance. In terms of data transfer performance, the server's PCIe is of version 3.0, with a transfer rate close to 8GB/s. This is comparable to the maximum transfer rate between the host and the device.

\begin{figure*}[ht]
\centering
\includegraphics[width=0.7\textwidth]{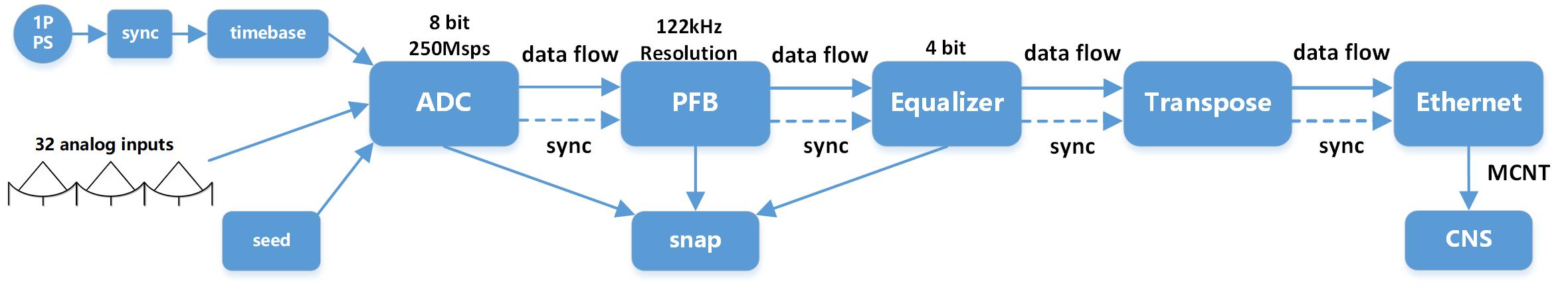}
\caption{Data flow block diagram of each F-engine. }
\label{fig:F-engine}
\end{figure*}

\subsection{F-engine} \label{subsec:F-engine}
The diagram of the F-engine module is shown in Figure \ref{fig:F-engine}. The Tianlai cylinder correlator system is an improvement upon the Tianlai dish correlator system, with enhancements including an increased number of input signals and additional new functions. The Tianlai dish correlator is very similar to the PAPER experiment correlator, which also uses the ROACH2 system.  Please refer to \citep{niuchenhui2019-roach} for details. Here, we provide a concise overview of the F-engine's process and the functionality of the CASPER yellow block.

Each ROACH2 board is connected with two ADC boards through Z-DOK+ connectors. The ADC board is the adc16x250-8 coax rev2 Q2 2013 version, which uses 4 HMCAD1511 chips and provides a total of 16 inputs. It samples 16 analog signal inputs with 8 bits at a rate of 250 Msps. 

The output digital signal of the \texttt{adc16x250} block is Fix\_8\_7 format, which indicates an 8-bit number with 7 bits after the decimal point. The ADC chip is accompanied by a control program developed by David MacMahon from the University of California, Berkeley. This program is responsible for activating the ADC, selecting the amplification level, calibrating the FPGA input delay, aligning the FPGA SERDES blocks until data is correctly framed, and performing other related tasks. A comprehensive user's guide for the ADC16 chip is accessible on the CASPER website\footnote{https://casper.astro.berkeley.edu/wiki/ADC16x250-8\_coax\_rev\_2}. According to the actual range of input signal power, we conducted linearity tests on the ADC at various gain coefficients and also assessed the linearity of the correlator system. Ultimately, the ADC gain coefficient was set to 2.

The analog-to-digital converted data from the ADC is transmitted to the PFB (Polyphase Filter Bank) function module. PFB is a computationally efficient implementation of a filter bank, constructed by using an FFT (Fast Fourier Transform) preceded by a prototype polyphase FIR filter frontend \citep{PFB}. The PFB not only ensures a relatively flat response across the channels but also provides excellent suppression of out-of-band signals. The PFB is implemented using the models \texttt{pfb\_fir} and \texttt{fft\_biplex\_real\_2x} from the CASPER module library.
 
Each \texttt{pfb\_fir}\footnote{https://casper.astro.berkeley.edu/wiki/Block\_Documentation} block (the signal processing blocks mentioned in this article can all be linked to the detailed page from here) processes two signals, configured with parameters including a PFB size of $2^{11}$, a Hamming window function, four taps, input width of 8 bits, an output width of 18 bits, and other settings. Each block takes two input signals, and a total of 16 \texttt{pfb\_fir} blocks are used to process 32 input signals. Each \texttt{fft\_biplex\_real\_2x} block processes four input data streams and outputs two sets of frequency domain data. configured with parameters including an FFT size of $2^{11}$, an input width of 18 bits, an output width of 36 bits, and other settings. The parameter settings are based on the scientific requirements of the Tianlai project, which calls for a signal resolution of less than or equal to 0.2 MHz. There are eight \texttt{fft\_biplex\_real\_2x} blocks, with each block taking in four data streams and outputting two sets of frequency domain data. The PFB module is flexible, making it very easy to adjust the parameters according to one's requirements, such as the FFT size, PFB size, and the number of taps in the CASPER block.

The data output of the PFB module is 36 bits, which essentially represents a complex number with 18 bits for the real part and 18 bits for the imaginary part. Considering factors such as data transmission and hardware resources, the data is usually effectively truncated. In our case, we will truncate the complex number to have a 4-bit real part and a 4-bit imaginary part. Prior to quantizing to 4 bits, the PFB output values pass through a scaling (i.e. gain) stage. Each frequency channel of each input has its own scaling factor. The purpose of the scaling stage is to equalize the passband before quantization, so this stage is often referred to as EQ. The scaling factors are also known as EQ\footnote{https://casper.astro.berkeley.edu/wiki/PAPER\_Correlator\_EQ} coefficients and are stored in shared BRAMs. 

The quantized data cannot be sent directly to the X-engine. Before sending it, we divide the frequency band and sort the data in a format that facilitates the relevant calculations. This module is called Transpose, and it is divided into four submodules. Each submodule processes 1/4 of the frequency band, resulting in a total of 256 frequency channels. The number of submodules corresponds to the number of 10\,GbE network interface controllers (NICs) on the ROACH2 board, with each NIC used to receive and send data from the output of a transpose submodule. This module performs the data transpose, also known as a ``corner turn'' to arrange the data in the desired sequence. Additionally, it is responsible for generating the packet headers, which consist of \texttt{MCNT} (master counter), \texttt{Fid} (F-engine id), and \texttt{Xid} (X-engine id). The current parameter configuration of the sub-module is tailored for scenarios with 256 inputs or fewer. However, David MacMahon, the researcher behind the PAPER correlator system, has included sufficient spare bits in the design, enabling the adjustment of model parameters based on specific input conditions and accommodating scalability and additional use cases.

The data is already in a form that is easy for X-engine to compute, we want to send it to X-engine, so the data comes to the Ethernet module. It contains four sub-modules and receives data from four transpose sub-modules. Each submodule has a \texttt{Ten\_GbE\_v2} block, where we can set the MAC address, IP address, destination port and other parameters using Python or Ruby script. 

\begin{figure*}[ht!]
\centering
\graphicspath{{./fig/}}
\includegraphics[width=0.65\textwidth]{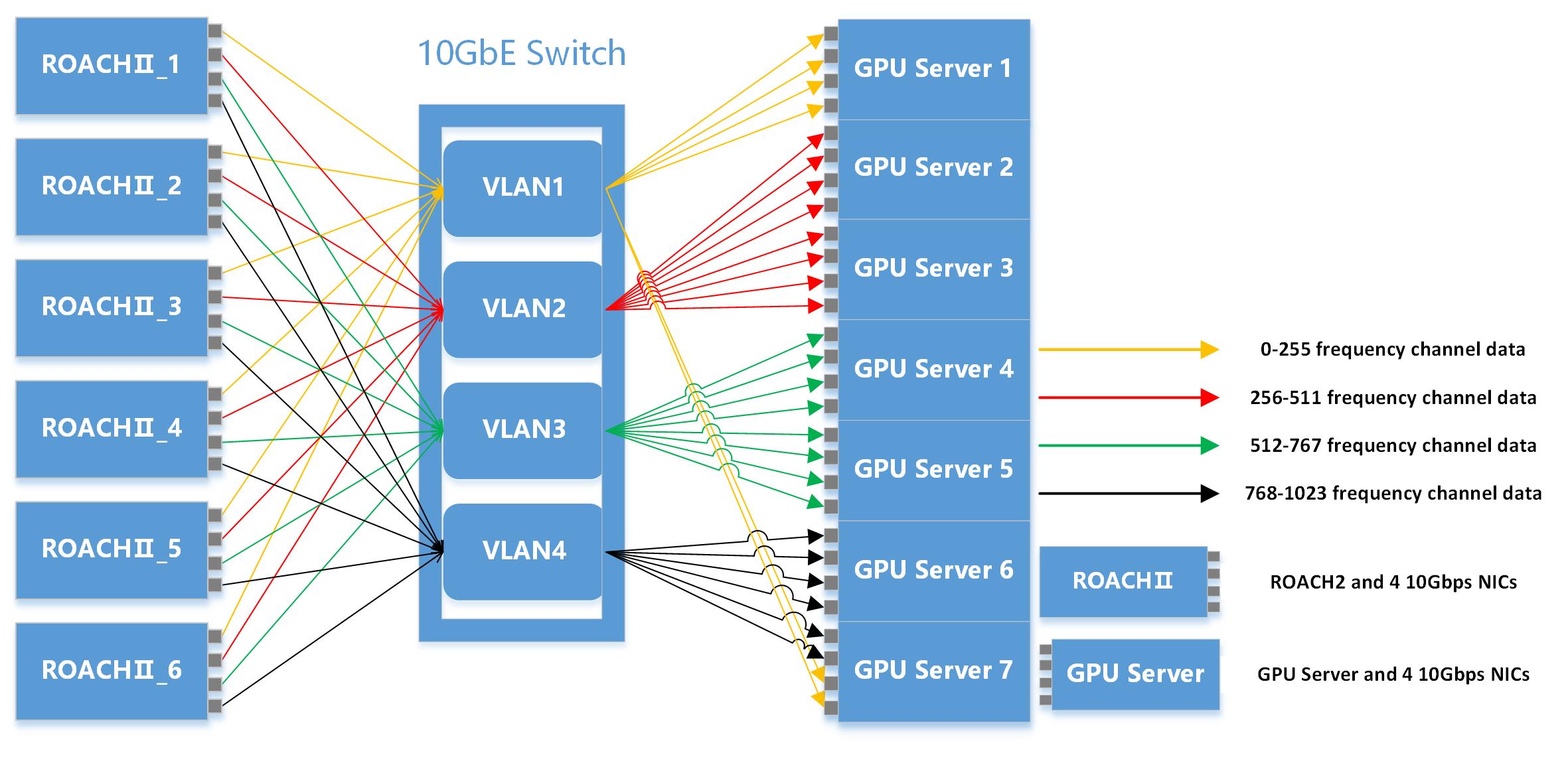}
\caption{Diagram of data transmission between X-engine and F-engine. Each ROACH2 board has 4 10\,GbE ports, and each port transmits data from 256 frequency channels. The 10\,GbE switch is configured with 4 VLANs (Virtual Local Area Networks), which offers benefits in simplicity, security, traffic management, and economy. Each VLAN receives UDP data from 256 frequency channels and sends them to the GPU servers. 
\label{fig:switch}}
\end{figure*}

\subsection{Network} \label{subsec:Network}

The data of the F-engine module is sent out through the ROACH2 network port and transmitted to the network port of the target GPU server through the 10\,GbE switch. The network transmission model of the correlator system is dependent on the bandwidth of a single frequency channel and the number of frequency channels calculated by the GPU server. The diagram of data transfer from F-engine to X-engine is shown in Figure \ref{fig:switch}.

The frequency domain data in F-engine has a total of 1024 frequency channels. Given the 250 Msps sampling rate, each frequency channel has a width of $\Delta \nu$ = 125 / 1024 MHz = 122.07 kHz. Each GPU node processes 128 frequency channels with a bandwidth of 128 $\times$ 122.07 kHz = 15.625 MHz. The number of frequency channels processed by the GPU server is determined by hashpipe.

\begin{figure}[ht!]
\graphicspath{{./fig/}}
\plotone{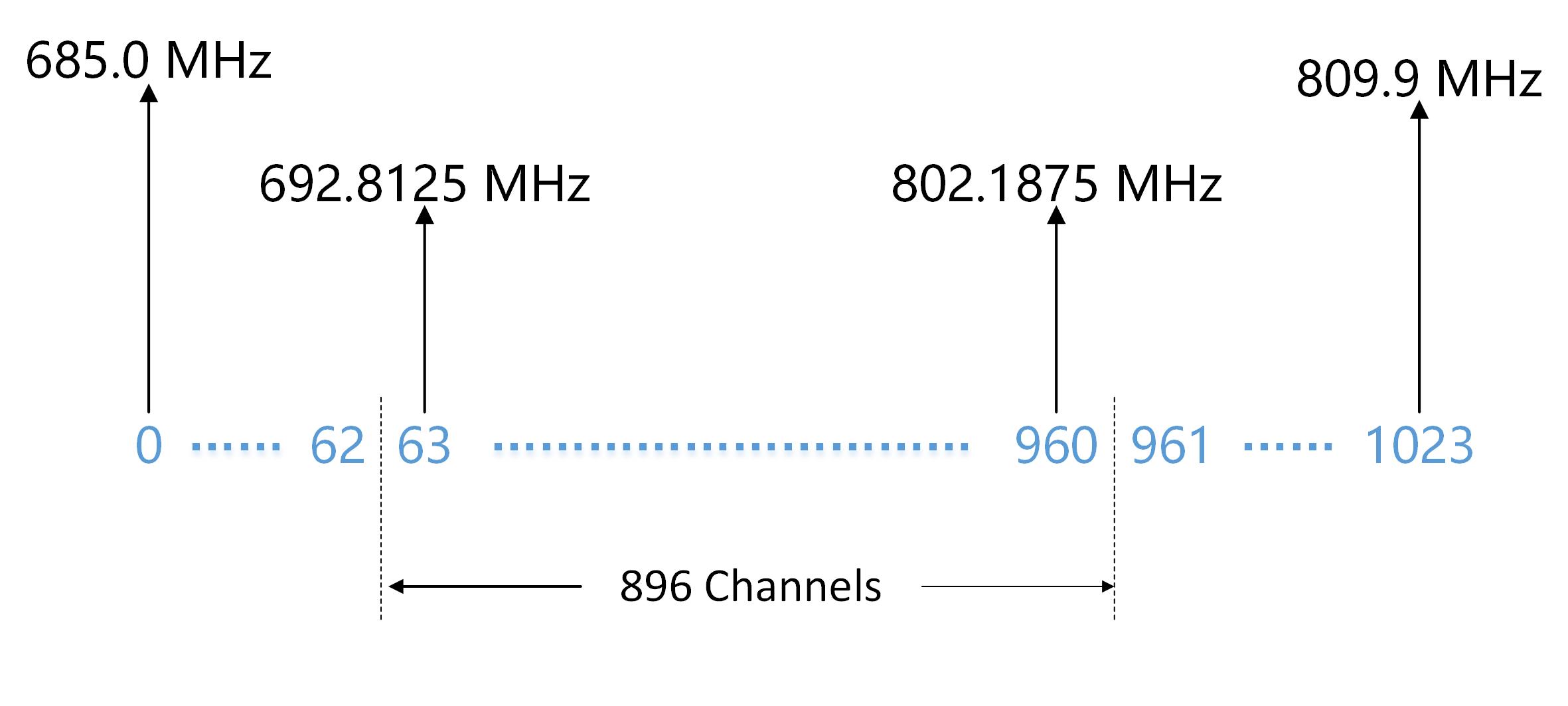}
\caption{The relationship between the FFT channels and the radio frequency. There are a total of 1024 frequency channels, and the input signal's effective frequency range is 700-800 MHz. The correlator's actual processing frequency range is 692.8125-802.1875 MHz, which includes a total of 896 frequency channels.
\label{fig:fft_range}}
\end{figure}

The analog part of the Tianlai digital signal processing system uses replaceable bandpass filters, with the bandpass set to 700 MHz $\sim$ 800 MHz. We have chosen to utilize seven GPU nodes to implement the X-engine component. These seven GPU nodes process data for the central 896 frequency channels, covering a bandwidth of approximately 109.375 MHz from 692.8125 MHz to 802.1875 MHz, as shown in Figure \ref{fig:fft_range}. The final GPU node is dedicated to receiving data from the first 32 and last 32 frequency channels out of the 896 frequency channels.

The data transfer rate of a single network port of the ROACH2 board is 8.0152 Gbps, so the total data transfer rate of 6 ROACH2 boards is 6 $\times$ 4 ports $\times$ 8.0152 Gbps = 192.3648 Gbps. The data reception rate of a single network port on the GPU node is 192.3648 Gbps / (8 $\times$ 4) = 6.0114 Gbps. At present, the number of input signals for the correlator system ranges from 32 to 256. While our correlator system is designed for 192 input signals, we conducted data transmission simulations with 256 input signals. Under these conditions, the data reception rate of a single network port on the GPU node stands at 8.0152 Gbps.

In our system, each GPU server has four 10\,GbE ports. For the Tianlai cylinder correlator system, we require a total of 6 ROACH2 boards $\times$ 4 ports + 7 GPU servers $\times$ 4 ports = 52 ports on a 10\,GbE switch. So we selected the Mellanox SX1024 switch which has 48 ports of 10\,GbE and 12 ports of 40\,GbE. Ports 59 and 60 on the switch can be subdivided into four 10\,GbE ports, providing ample capacity for our application.


The transpose module is designed with extra bits reserved in the blocks related to the parameter \texttt{fid}. The number of bits in the \texttt{fid} parameter is directly linked to the maximum number of F-engines in the correlator system. By utilizing these additional bits, the correlator can be configured to accommodate a greater number of input signals. In terms of the F-engine, theoretically, there could be an infinite number of input channels, and the number of ROACH2 boards can be increased based on the input channel number. The capacity of X-engine determines the upper limit of input channels, depending on the processing capacity of the GPU servers for a single frequency point. Since each frequency point should contain all the input channel information, the processing capability of the GPU servers for a single frequency channel affects maximum number of input channels. Currently, a single server theoretically has the capability to handle over 20,000 input channels if it only processes one frequency channel. However, this may need an extremely large-scale switch network. 

The relationship between the number of input channels and the output data rate is as follows:
\begin{equation}
\frac{1}{2} \times N(N+1) \times f\_ch \times 2 \times f\_b / Integration\_time
\label{eq:vis}
\end{equation}
where $N$ represents the number of input channels, $f\_ch$ represents the number of frequency channels, $f\_b$ represents the number of bytes in a single frequency channel. The multiplying factor 2 is because frequency channels are complex numbers. 

\subsection{X-engine} \label{subsec:X-engine}

The primary role of the X-engine is to perform cross-correlation calculations. The X-engine receives the data from the F-engine in packets, which are then delivered to different computing servers, where the conjugate multiplication and accumulation (CMAC) are done. The hardware for this part consists mainly of six Supermicro servers and one Dell server. We list the main equipment of the X-engine in Table \ref{table:servers}.

\begin{table}[ht]
\centering
\begin{tabular}{ccc}
\hline
                 & Supermicro (4U)                                                                   & Dell (2U) \\ \hline
PCIe             & 3.0                                                                                  & 4.0                     \\
Graphics Card    & Dual GTX 690                                                                  & One RTX 3080            \\
CPU              & Dual Intel E5-2670                                                                   & Dual Intel E5-2699          \\
NIC              & Dual 2-port 10\,GbE                                                                   & Dual 2-port 10\,GbE\\
Memory           & 128 GB RAM                                                                           & 256 GB RAM        \\
OS               & Centos7                                                                              & Rocky8                  \\ \hline
\end{tabular}
\caption{List of X-engine equipment}
\small
\label{table:servers}
\end{table}

\begin{figure*}[ht!]
\centering
\graphicspath{{./fig/}}
\includegraphics[width=0.55\textwidth]{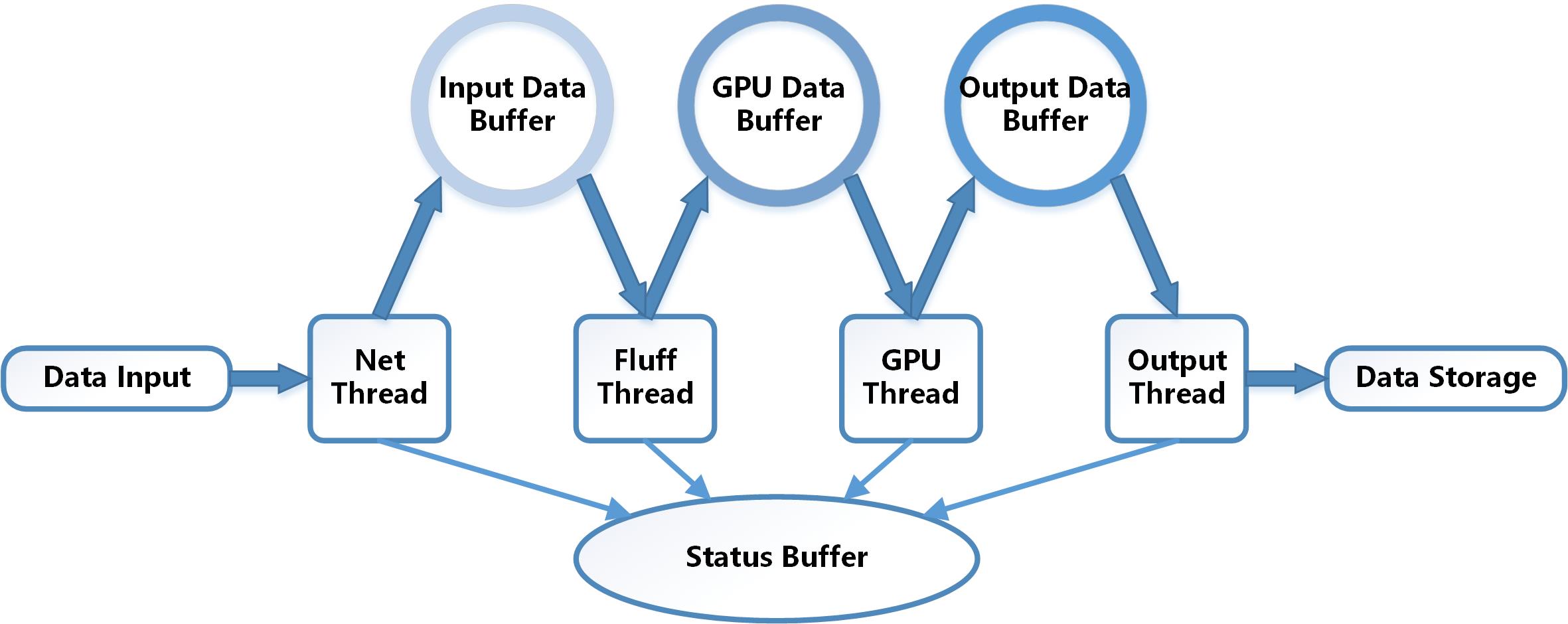}
\caption{Hashpipe thread manager diagram.
\label{fig:hashpipe}}
\end{figure*}

The X-engine part consists of seven GPU nodes. To ensure that they integrate the data at exactly the same time duration, they must be synchronized together. A script has been developed to achieve this, and its basic procedure is as follows. 
First, initialize the hashpipes of 7 GPU nodes; Second, start the hashpipe program of the first GPU node; 
Third, read out the MCNT value in the current packet and calculate a future (several seconds later) MCNT value to act as the aligning time point. 
Finally, all GPU nodes work simultaneously when their hashpipe threads receive a packet contains the calculated aligning MCNT value. 

The data operation in the X-engine is managed by the hashpipe software running on CPU and GPU heterogeneous servers.
Hashpipe was originally developed as an efficient shared pipe engine for the National Astronomical Observatory, the Universal Green Bank Astrospectrograph \citep{green_bank}. It was later adapted by David MacMahon of U.C. Berkeley, it can be used for FX correlators \citep{PAPER_correlator}, pulsar observations \citep{peixin}, Fast Radio Bursts detection \citep{dish_frb} and the search for extraterrestrial civilizations \citep{breakthrough_listen}. The core of the hashpipe is the flexible ring buffer. It simulates contiguous memory blocks, realizes data transmission and sharing among multiple threads, and uses the central processing unit to control startup and shutdown, etc. The ring buffer is used to temporarily store and deliver the data packets to ensure that the data is captured quickly and distributed in the correct order. 

Each hashpipe instance in our system has a total of four threads and three buffers, as shown in Figure \ref{fig:hashpipe}. To process the four 10\,GbE ports data stream, four hashpipe instances are created. 
In each instance, the basic data process can be concluded as follows. First, \texttt{net\_thread} receives the packets from the GPU server's 10\,GbE port. According to the packet format, the valid data is extracted and the packet header is analyzed. Packets are time-stamped, and if they arrive at the GPU server out of order, they can be rearranged into the appropriate time series and written to the input data buffer, which is passed onto the next thread once a consecutive block of data is filled. The \texttt{fluff} thread ``fluffing'' the data, fluffs 4bit+4bit complex data into 8bit+8bit complex data in the thread. The data is ``fluffed'' and temporarily stored in the GPU input data buffer until it is fetched by \texttt{gpu\_thread}. Then \texttt{gpu\_thread} transfers the data to the graphics processor to perform complex calculations and then writes the results to the output data buffer. The CMAC process uses the xGPU \footnote{https://github.com/GPU-correlators/xGPU}\citep{xgpu}, which is written in CUDA-C and is optimized on GPU memory resources by specific thread tasks. The cross-correlation algorithm involves computing the cross-power spectrum at a specific frequency observed by a pair of stations, known as a baseline. By processing a sufficient number of baselines, a detailed power spectrum representation can be derived, enabling the generation of an image of the sky through an inverse Fourier transform in the spatial domain. The algorithm's implementation on Nvidia's Fermi architecture sustains high performance by utilizing a software-managed cache, a multi-level tiling strategy, and efficient data streaming over the PCIe bus, showcasing significant advancements over previous GPU implementations. The \texttt{output} thread gets the data from the output data buffer and transmits it to the storage server through the switch. Hashpipe provides a status buffer that extract key-value pairs in each thread. This key value is updated every running cycle. The status can be viewed using a GUI monitor that has been written in both Python and Ruby.

\subsection{Data storage} \label{subsec:data storage}

At the beginning of the design, two schemes for data storage were considered. One is that the data is stored on each GPU server, and it is read and combined when used. Due to the large number of GPU servers, this method is too cumbersome. The other is that the data is transmitted from each GPU server to the master computer in real-time, and the data is stored in the master computer. This method is convenient for data use and processing, so the second scheme is adopted.

Each GPU node has 4 hashpipe instances, and the \texttt{output} thread of each hashpipe instance sends data to a dedicated destination port. A total of 28 different UDP ports are used for the 7 GPU servers. The data acquisition script, written in Python, collects data from all 28 UDP ports and combines them . Currently, the integration time is set to approximately 4 seconds, resulting in a data rate of about 150 Mbps for each network port. The total data rate for all seven servers with 28 ports amounts to approximately 4.2 Gbps. Therefore, a 10\,GbE network is capable of the transmission of the data. Therefore, a 10\,GbE network is capable of handling the data transmission. Finally, the data are saved onto hard drives in the HDF5 format. Additional information such as integration time, observation time, telescope details, and observer information is also automatically saved in the file.

\subsection{CNS control module} \label{subsec:Noise Source Module}

\begin{figure}[ht!]
\centering
\graphicspath{{./fig/}}
\plotone{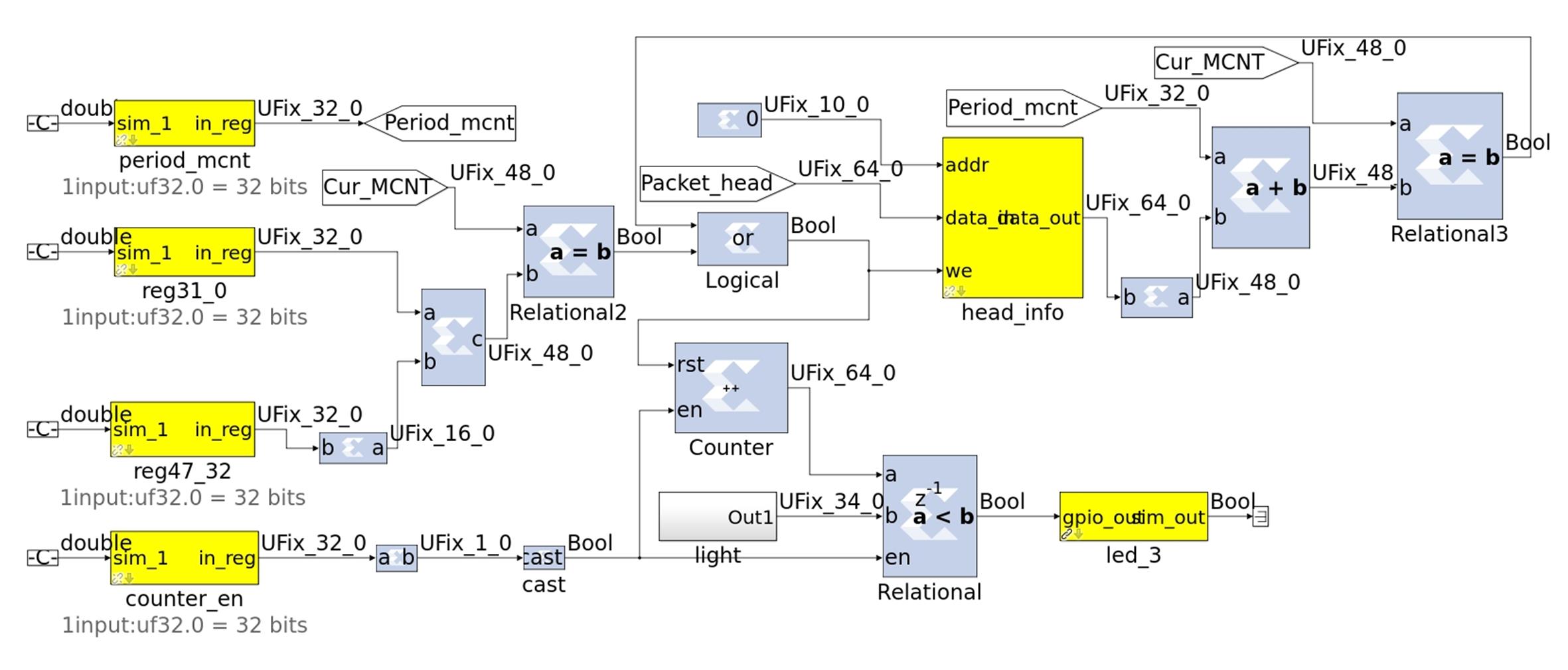}

(a)

\plotone{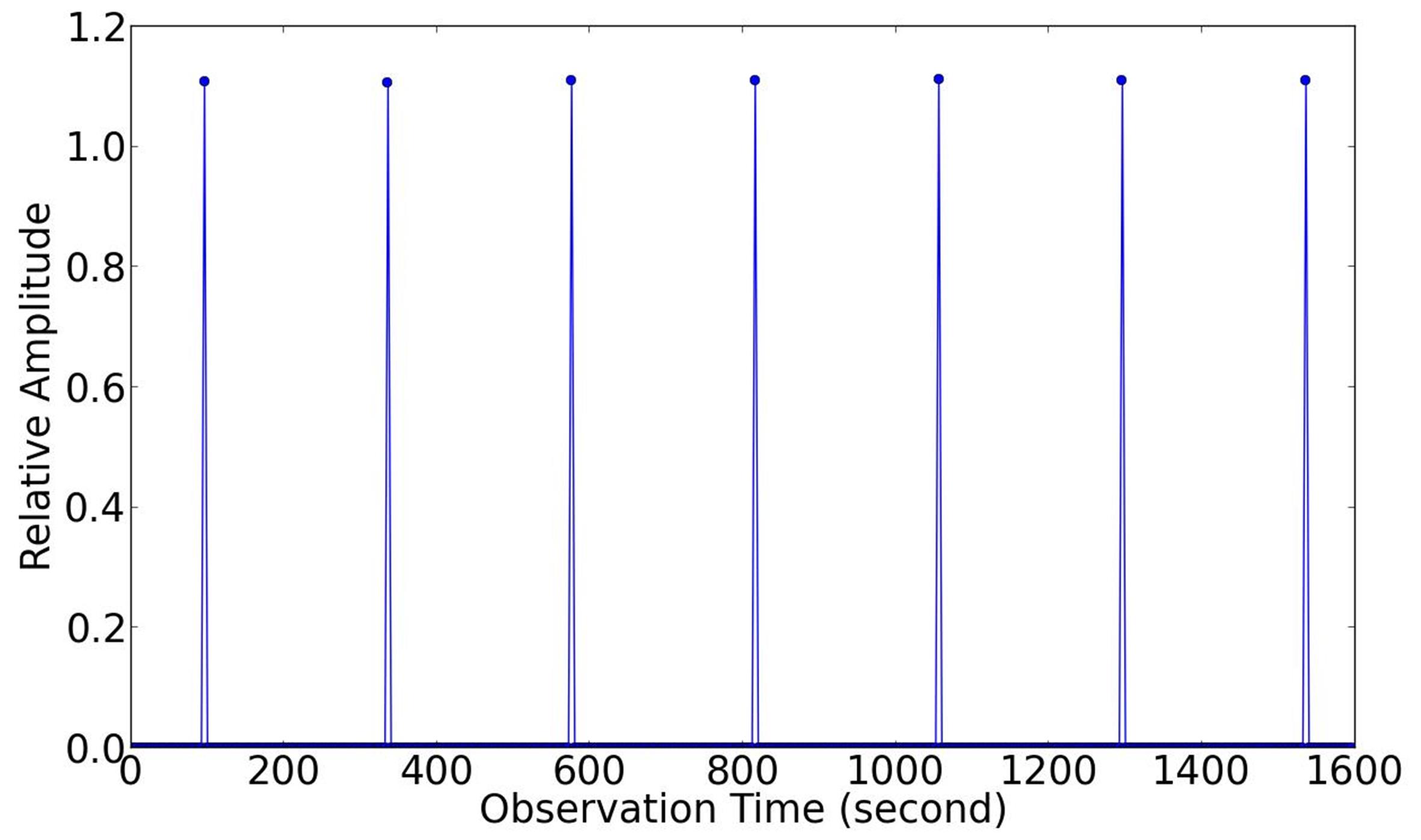}

(b)
\caption{(a) Design of the calibrator noise source control module. (b) The periodic CNS signal in the cross-correlation results is aligned with  the integration time.
\label{fig:noise_source}}
\end{figure}

During the drift scan observation of the Tianlai cylinder array, the system needs to be calibrated by a calibrator noise source (CNS). The CNS periodically broadcasts a broadband white noise of stable magnitude from a fixed position, so the system gain can be recovered \citep{tlpipe,zsf_eigen}. One requirement in the data processing part is to let the CNS's signal fall exactly in one integration time interval, so it needs to be aligned to the integration time. To achieve this, a logical ON/OFF signal from the cylinder correlator is necessary. In order to meet this requirement, We have introduced a noise source control function to the correlator system. This control function is implemented through the \texttt{noise\_source\_control} block in the F-engine, as shown in Figure \ref{fig:noise_source}(a). 

First, the script enables \texttt{counter\_en} block to initialize the module. Second, the hashpipe instance on the GPU node returns the MCNT value of its current packet. The script uses this value to calculate the  CNS MCNT value (an MCNT value at a future time, when the MCNT value in the F-engine is equal to this value, the CNS is turned ON) and sets that CNS MCNT to \texttt{reg31\_0} block and \texttt{reg47\_32} block. Third, the CNS on/off period is converted to the change value of MCNT and set \texttt{period\_mcnt} block to this value. Fourth, set the GPIO's working time to \texttt{light} block, which is on the ROACH2 board.
Finally, the GPIO periodically sends out a logical signal to turn the CNS on or off.

We tested the accuracy of the CNS control module and its actual output result, as shown in Figure \ref{fig:noise_source}(b). The CNS is activated based on a pre-set MCNT value and is aligned precisely with the integration time interval.

\section {Testing and Experimentation} \label{sec:test}


\subsection{ADC testing} \label{sec:adc test}

\begin{figure}[ht!]
\centering
\graphicspath{{./fig/}}
\plotone{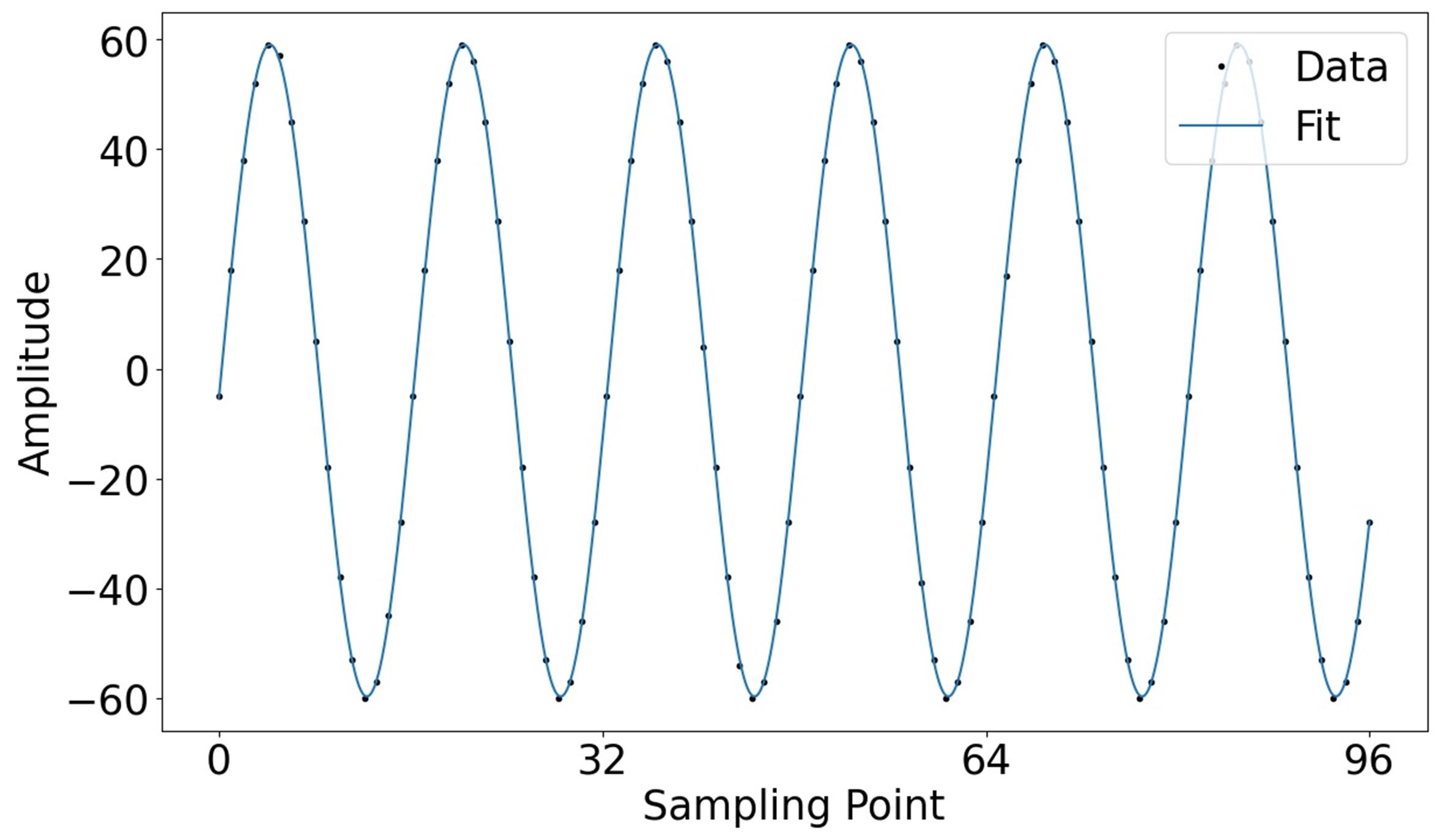}

(a)

\plotone{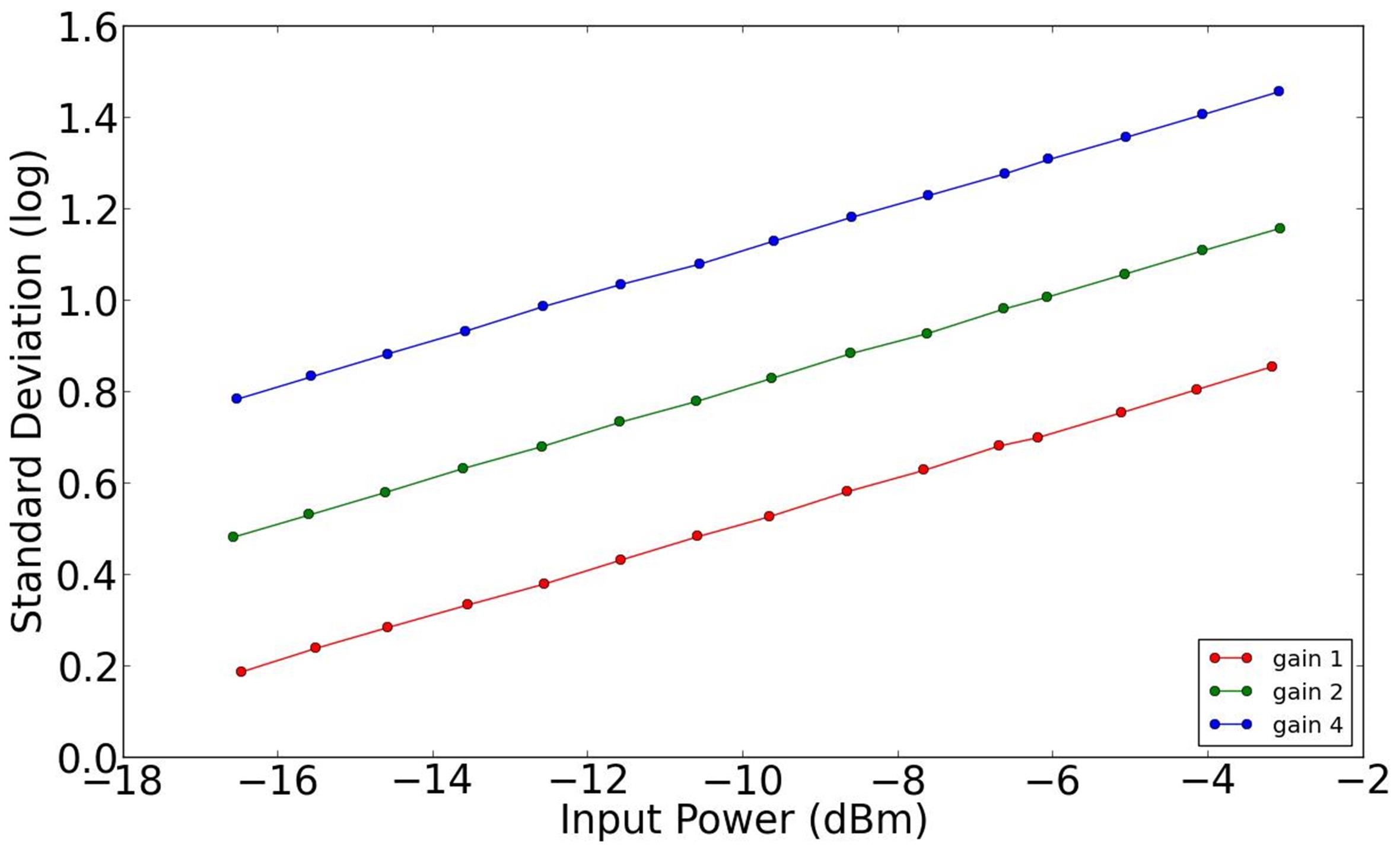}

(b)
\caption{(a) The ADC output data points and the fitting curve. 
(b) the logarithm of the standard deviation of the ADC output with gain coefficient 1, 2 and 4 was plotted at different input power levels. 
\label{fig:ADC_test}}
\end{figure}

The importance of ADCs lies in their quality and performance, as these factors bear a direct impact on the overall functionality of the systems they inhabit.
To verify the sampling correctness of the ADC, we input a 15.625 MHz sinusoidal wave signal into the ADC and fit the digitized data. The sampling points and fitting result are shown in Figure \ref{fig:ADC_test}(a).
The correlator system requires the ADCs to have linearly sampled output at different signal levels. 
We plot the logarithm of the standard deviation of the ADC output with three different gain coefficients as a function of different input power levels, and the results are shown in Figure \ref{fig:ADC_test}(b). No obvious nonlinearity is found in the testing power range.

\subsection{Phase testing}
\label{subsec:Phase}

We verify the phase of the visibility (cross-correlation result) by two input signals, whose phase difference is determined by a cable length difference. We use a noise source generator to output the white noise signal and the signal is divided into two ways by a power splitter. Then, the two signals are fed into the ROACH2 board through two radio cables of different lengths. The cable length difference is 15m. The two signals can be depicted as $S_1=A_1 e^{i(2\pi ft+\phi_0)}$ and $S_2=A_2 e^{i(2\pi f(t+\tau) + \phi_0 )}$, where $A$ is the wave amplitude, 
$\phi_0$ is an arbitrary initial phase, $f$ is frequency, $\tau$ is the delay incurred by the unequal-length RF cables. The visibility of two signals is 
\begin{equation}
V = <S_1^\ast \cdot S_2 > = A_1 A_2 e^{i2\pi f\tau} \label{eq:vis}
\end{equation}
The cable length difference of the two input signals is fixed, so the delay \texttt{$\tau$} is constant over time. As Eq. \ref{eq:vis} shows, the phase $\Phi = 2\pi\tau\cdot f$, $\Phi$ is a linear function of frequency, and the slope \textit{k} = 2$\pi$$\tau$. The delay \texttt{$\tau$} = $\Delta l/\tilde{c}$, where \textit{$\Delta$l} is the cable length difference, $\tilde{c}$ is the propagation speed of RF signal in coaxial cable. 

\begin{figure}[htbp]
 \centering
 \graphicspath{{./fig/}}
 \plotone{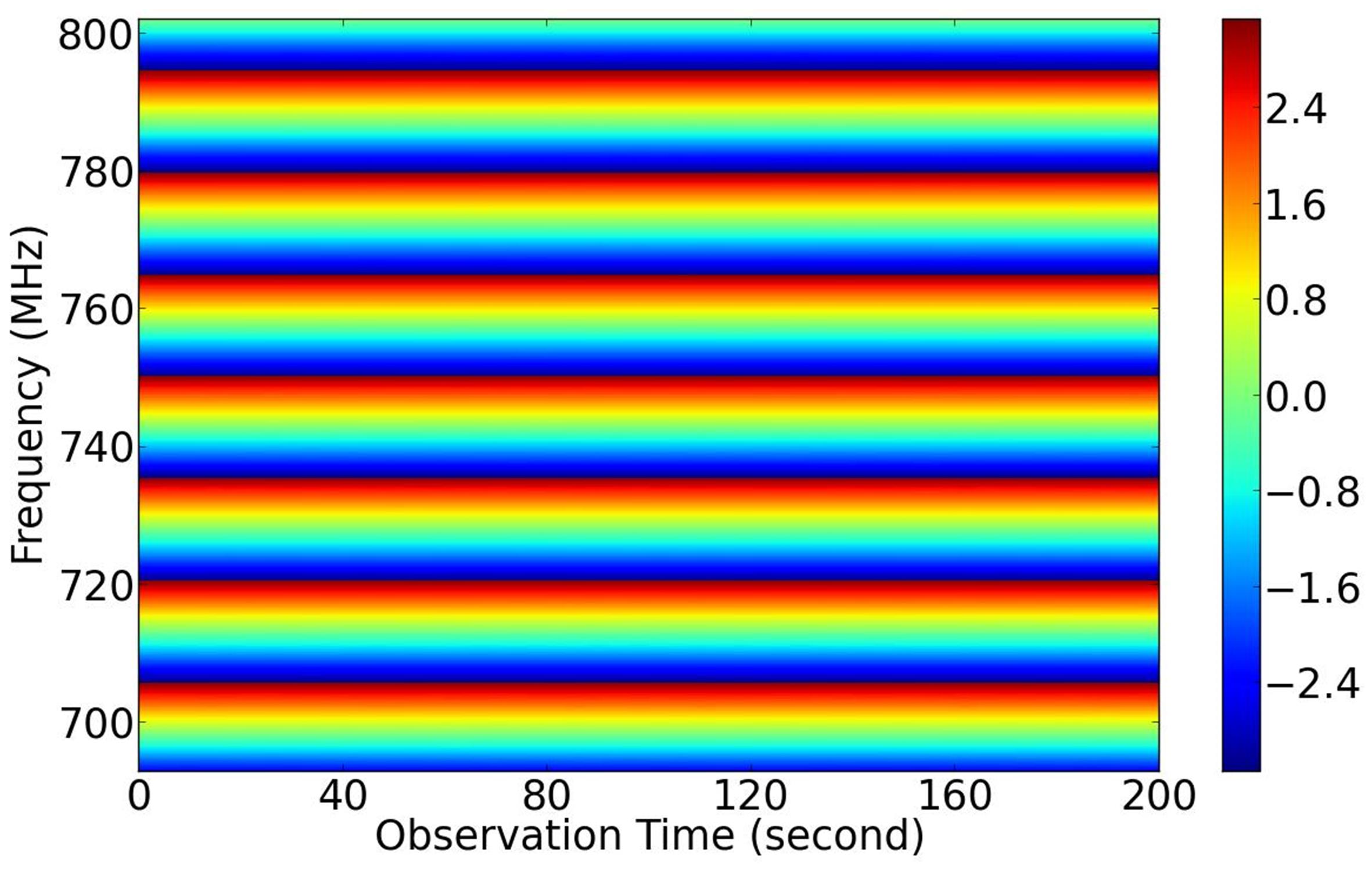}

 (a)

 \plotone{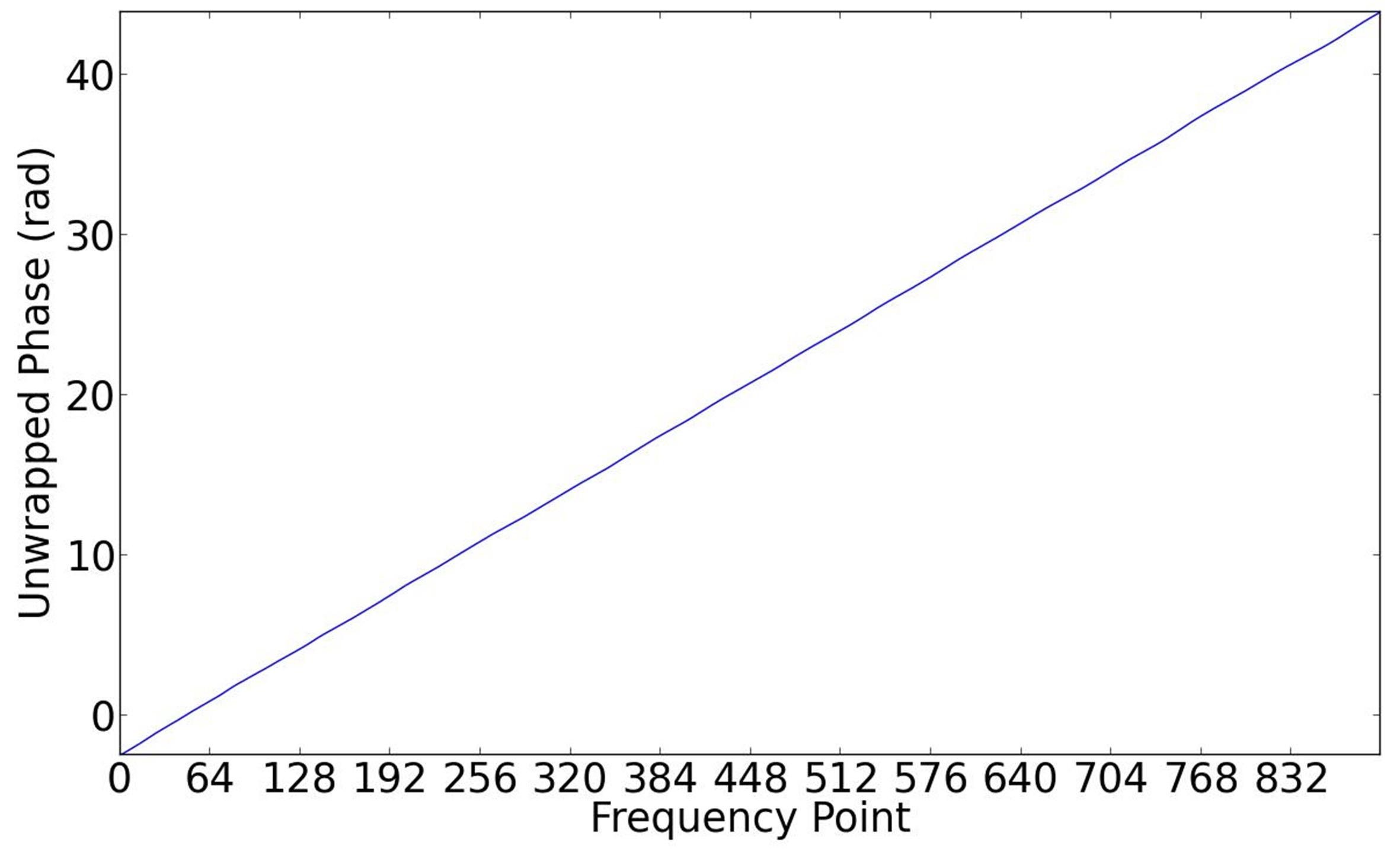}

 (b)
 \caption{
Phase correctness check of the correlator by two signals with fixed phase difference. The measured phase is consistent with the length difference of the two cables. 
}
 \label{fig:phase}
 \end{figure}

The measured waterfall 2D plot of phase of visibility output by our correlator in this experiment is plotted in Figure \ref{fig:phase}(a) and 1D plot (at one integration time) of phase as a function of frequency is shown in Figure \ref{fig:phase}(b). By calculating the curve slope in Figure \ref{fig:phase}(b), we obtain a propagation speed in the coaxial cable of about $0.78c$ (0.78 times speed of light in vacuum), which is consistent with the specification of the RF cable.

\subsection{Linearity of correlator system} \label{subsec:linearity}

\begin{figure}[htbp]
\centering
\graphicspath{{./fig/}}
\includegraphics[width=0.3\textwidth]{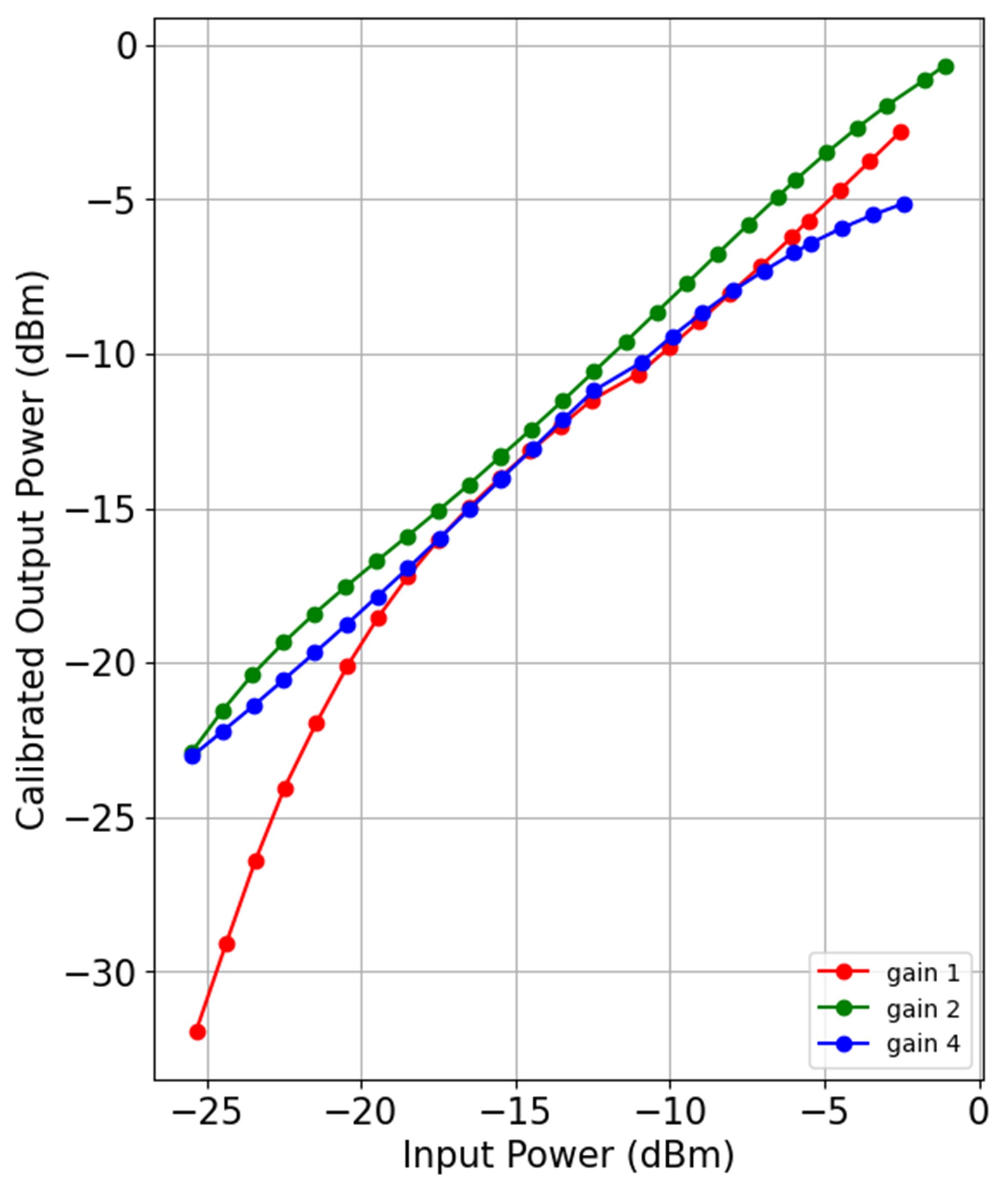}
\caption{Linearity of correlator system. The ADC gain coefficients of 1, 2, and 4 were used for system linearity testing. A gain coefficient of 2 was selected as the daily operational parameter value for the correlator.
\label{fig:linearity}}
\end{figure}

The linearity of our correlator is verified by comparing the input power levels and the output amplitudes. The results are shown in Figure \ref{fig:linearity}. Considering multiple factors, we have set the ADC gain coefficient for the correlator to 2. We can draw the conclusion that the linear dynamic range of our correlator is between -22 dBm to 0 dBm within the 125~MHz bandpass. 
In realistic observations, power levels output from the receivers vary 10 dB at most, so the 22 dB dynamic range of our correlator can satisfy our observation requirement.

\subsection{Sky observation} \label{subsec:observer}

\begin{figure}[ht!]
\graphicspath{{./fig/}}
\plotone{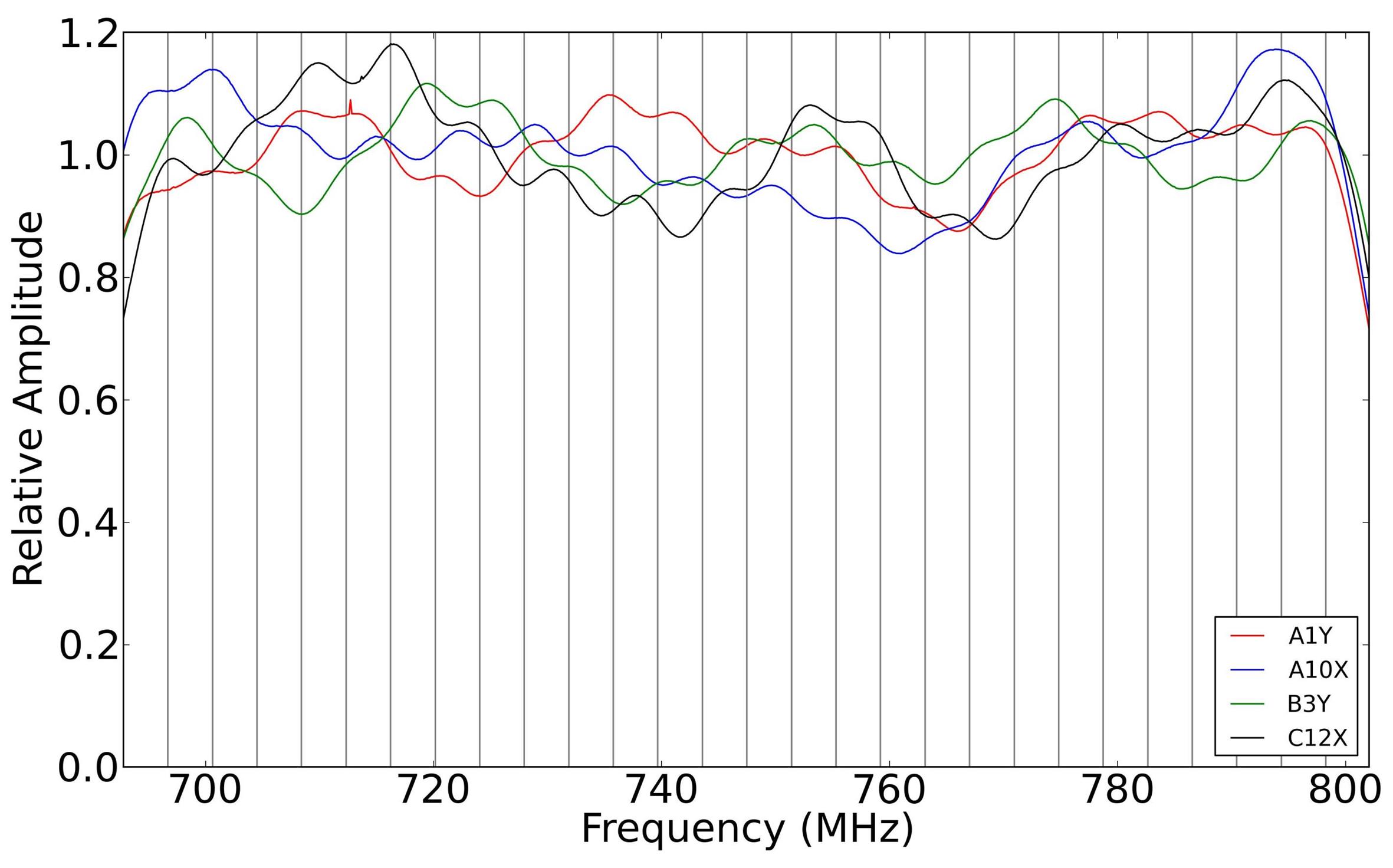}
\caption{Spectral response of feed A1Y, A10X, B3Y and C12X. 
\label{fig:spectrum}}
\end{figure}

The whole frequency band of each feed ranging from 692.8125 to 802.1875 MHz, is divided to 28 sub-bands. These sub-band have been sent to different hashpipe instances for correlation calculation. The final spectra are the combination of these 28 sub-bands. Some spectra of feeds (A10X, A19Y,
B27X and C12X) are plotted in Figure \ref{fig:spectrum}.
The Tianlai cylinder array is aligned in the N-S direction and consists of three adjacent cylinders. They are designated as A, B and C from east to west, and have 31, 32, and 33 feeds respectively. Each dual linear polarization feed generates two signal outputs. We use `X' to denote the output for the polarization along the N-S direction and `Y' along E-W direction. Spectra of the selected feeds in Figure \ref{fig:spectrum} are from three cylinders, and are smooth in adjacent frequency sub-bands. No obvious inconsistent processing amplitude in different sub-bands are found. 

In these spectra, a periodic fluctuation of about 6.8 MHz can be seen. They have been confirmed to result from the standing wave in the 15-meter feed cable \citep{jixia_sw}.

\begin{figure}[htbp]
\centering
\graphicspath{{./fig/}}
\plotone{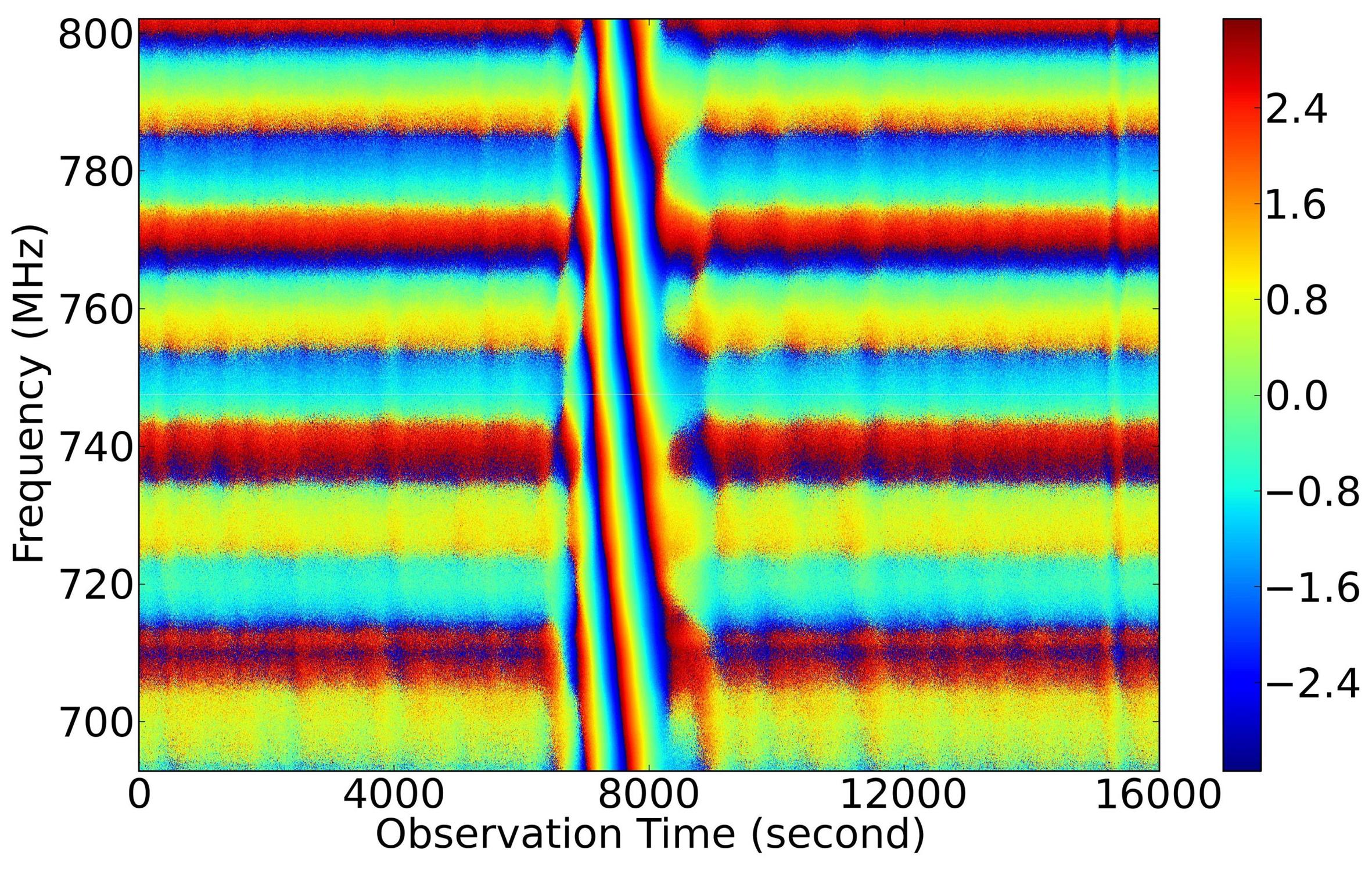}
\caption{Observational phase of Cassiopeia A at around 8000th second.
\label{fig:cygA}}
\end{figure}

 We made 4.4 hours (16000 seconds) of continuous observation since the night of Aug. 7th, 2023, and the data are shown in Figure \ref{fig:cygA}. The fringe of radio bright source Cassiopeia A occurred around  8000th second.
 
\begin{figure}[ht!]
\centering
\graphicspath{{./fig/}}
\includegraphics[width=0.3\textwidth]{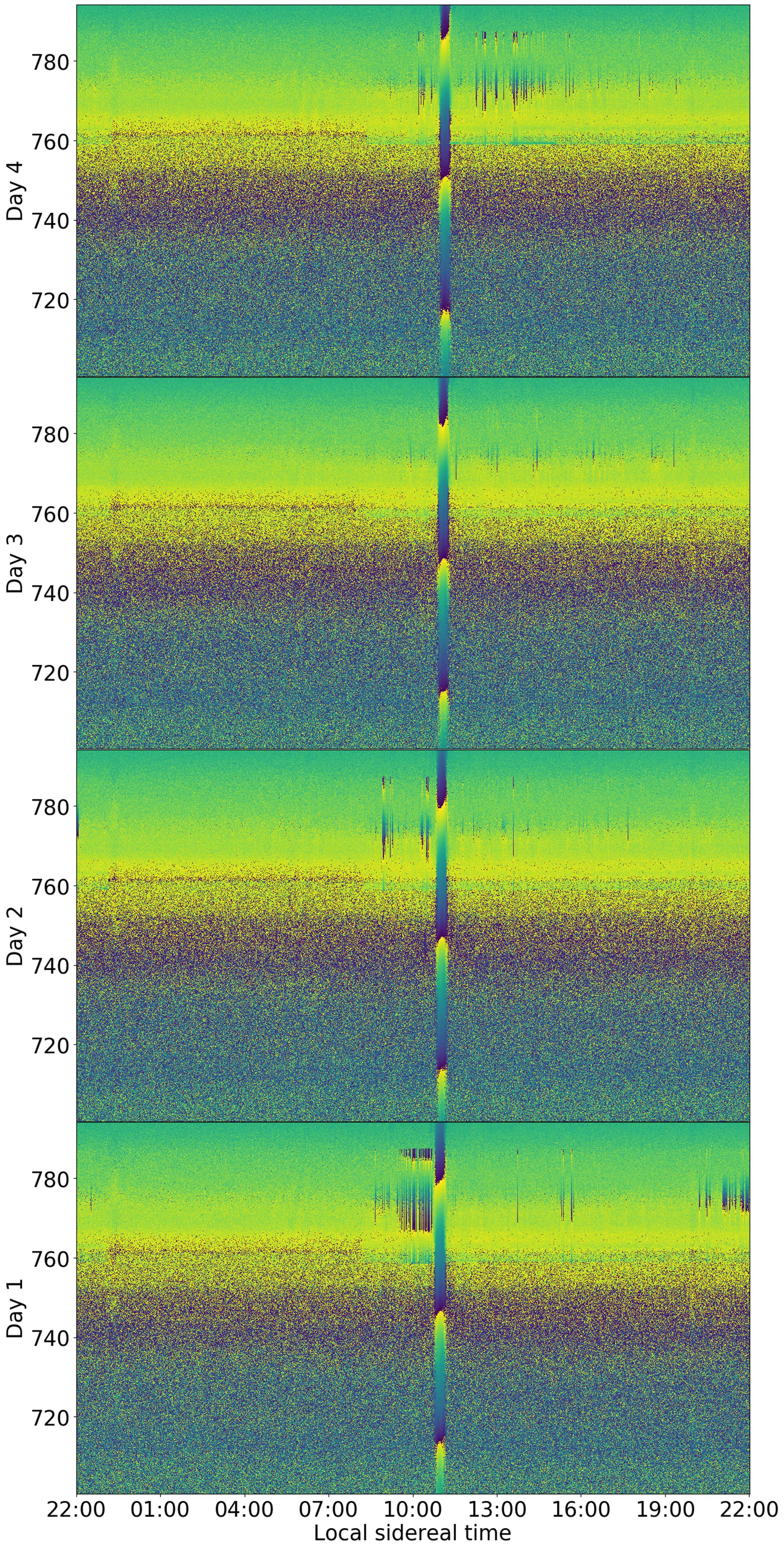}

(a)

\includegraphics[width=0.3\textwidth]{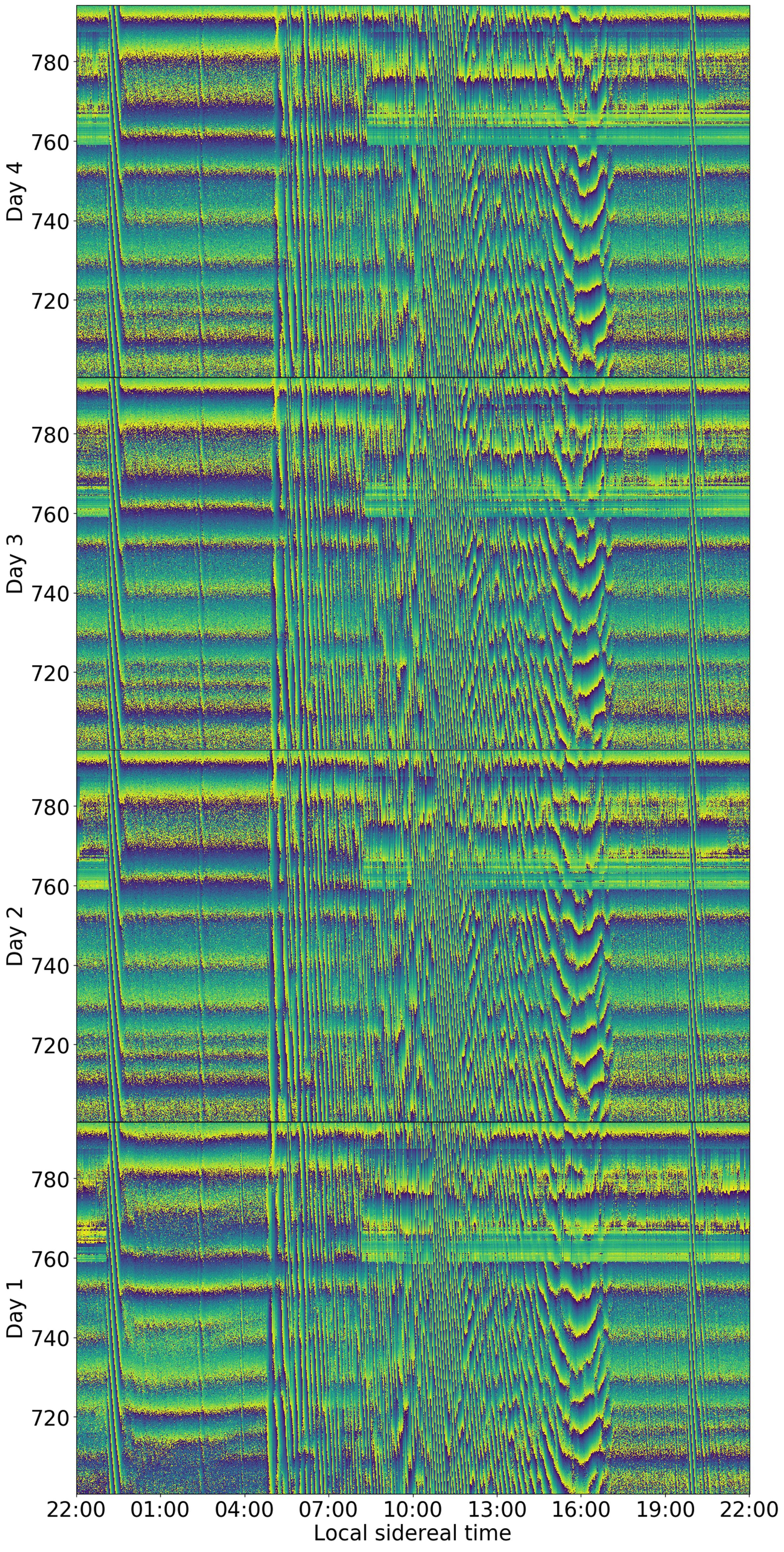}

(b)
\end{figure}
\begin{figure}[ht!]
\centering
\graphicspath{{./fig/}}
\includegraphics[width=0.3\textwidth]{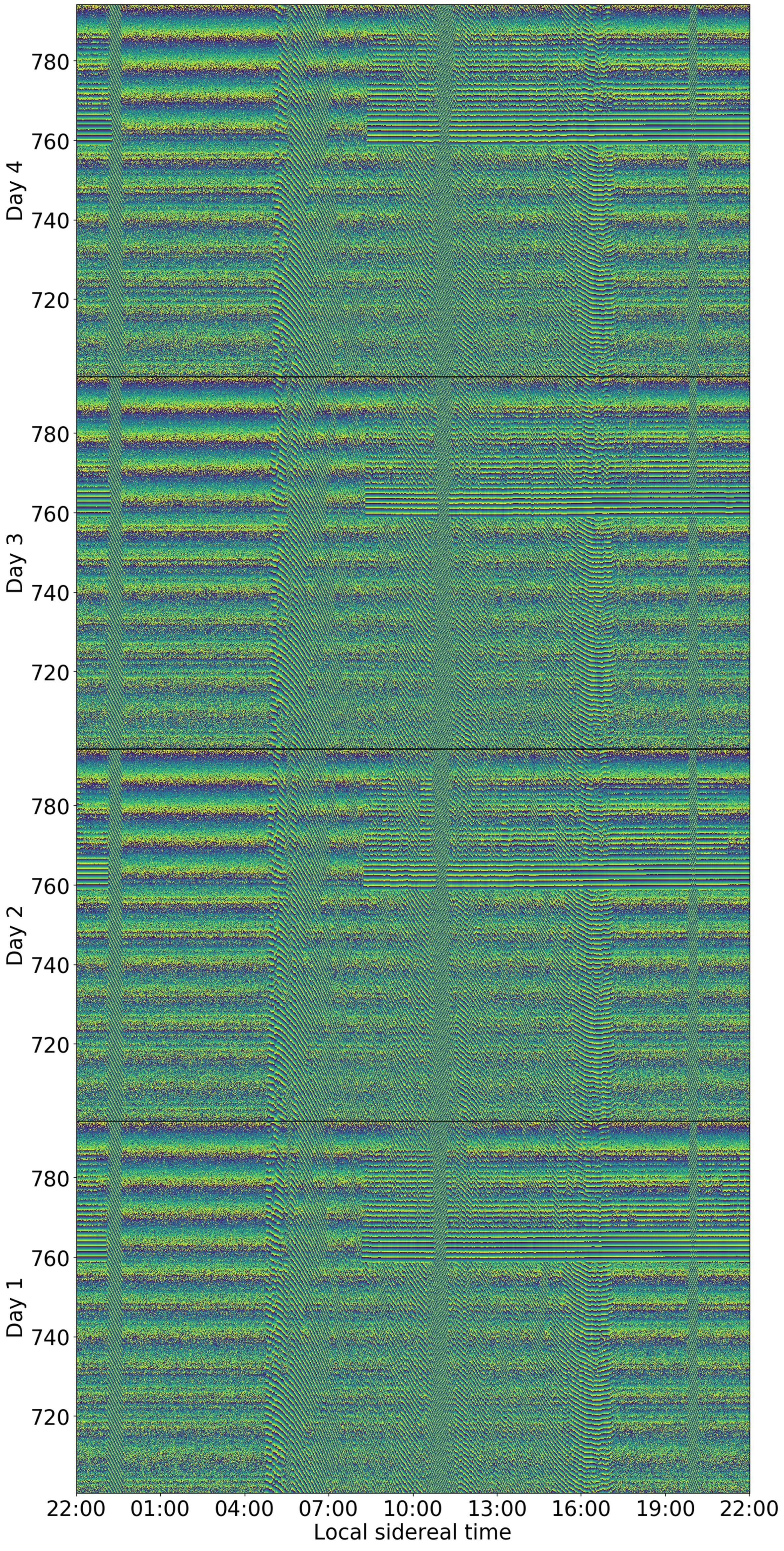}

(c)

\caption{Typical phase of raw visibilities as a function of LST and frequency for 4 days starting from Sept. 6th, 2023. (a) Baseline A3Y-A15Y; (b) Baseline A3Y-B18Y; (c) Baseline A3Y-C15Y.
\label{fig:4day_vis}}
\end{figure}

The continuous operation ability of the correlator is tested, and there is no fault in continuous operation for a month. We plot 4 days' continuous observation data of three baselines as a function of LST (Local sidereal time) and frequency, as shown in Figure \ref{fig:4day_vis}. The subplots from top to bottom show the baselines for two feeds (a) on the same cylinder, (b) on two adjacent cylinders, and (c) on two non-adjacent cylinders. Each subplot shows the result of four consecutive days starting from Sept. 6th, 2023; each day is a sub-panel from bottom to top.

\subsection{Power consumption} \label{subsec:Power}
All devices are powered by PDU (Power Distribution Unit),  and the voltage and current usage of the devices can be monitored through the PDU management interface. The entire correlator system uses a total of 3 PDUs. The six ROACH2 boards and the master computer are connected to one PDU. The first 7 GPU servers and the 10\,GbE switch are connected to another PDU. The last 7 servers and the 1\,GbE  Ethernet switch are connected to the third PDU.

The total power of the F-engine is 220 V $\times$ 3.5 A = 770 W, including six ROACH2 boards and one master computer. The total power of X-engine is 220 V $\times$ 17.5 A = 3850 W, including seven GPU servers, one 10\,GbE switch, and one 1\,GbE switch. Therefore, the total power of the whole correlator system is 770 W + 3850 W = 4620 W for 192 inputs. 
This is very energy-efficient for such a large-scale interferometer system.

\section {Summary} \label{sec:summary}

In this paper, the correlator is designed and deployed for the cylinder array with 192 inputs. Based on the basic hybrid structure of the ROACH2-GPU correlator, we have realized the data acquisition and pre-processing function by F-engine, which consists of six ROACH2 boards. The F-engine part is tested, debugged, and analyzed, works in the suitable linear range and the calibrator noise source is controlled in a cadence according to integration time. We conducted hardware testing and data storage design for the X-engine part and realized the complete and orderly data storage of 7 GPU servers. We use a DELL 2020 server, NVIDIA GeForce RTX3080 graphics card, and Rocky 8 system to achieve the X-engine function.

As Tianlai radio interferometric array is currently 
extending its scale, the correlator we design can increase the number of ROACH2 boards according to the number of input signals, and set the appropriate number of frequency points and the size of data packets. The X-engine part can use higher-level servers and graphics cards to combine multiple tasks and increase the work tasks of a single server to reduce the number of servers. Our future work is to implement it on larger systems.

\section*{Acknowledgements}
We acknowledge the support by the National SKA Program of China (Nos. 2022SKA0110100, 2022SKA0110101, and 2022SKA0130100), the National Natural Science Foundation of China (Nos. 12373033, 12203061, 12273070, 12303004, and 12203069), the CAS Interdisciplinary Innovation Team (JCTD-2019-05), the Foundation of Guizhou Provincial Education Department (KY(2023)059), and CAS Youth Interdisciplinary Team. This work is also supported by the office of the leading Group for Cyberspace Affairs, CAS (No.CAS-WX2023PY-0102) and CAS Project for Young Scientists in Basic Research (YSBR-063).

\bibliography{sample631}{}
\bibliographystyle{aasjournal}



\end{document}